\documentclass[aps,prd,reprint,superscriptaddress,nofootinbib]{revtex4-1}
\usepackage[all]{xy}
\usepackage{amsmath,amsthm,amssymb}
\usepackage[dvips]{graphicx}
\usepackage{comment}
\usepackage{array}
\usepackage{bm}
\allowdisplaybreaks[1]

\usepackage{color}
\usepackage{hyperref}

\newcommand{\prt}{\partial}
\newcommand{\qwe}{\begin{equation}}
\newcommand{\ewq}{\end{equation}}
\newcommand{\J}[1]{\left[#1\right]}
\newcommand{\tu}{\tilde{u}}
\newcommand{\tv}{\tilde{v}}

\date{\today}
\begin{document}

\title{Time-domain metric reconstruction for self-force applications}
\author{Leor Barack}
\author{Paco Giudice}

\affiliation{ Mathematical Sciences, University of Southampton, Southampton SO17 1BJ, United Kingdom}

\begin{abstract}
We present a new method for calculation of the gravitational self-force (GSF) in Kerr geometry, based on a time-domain reconstruction of the metric perturbation from curvature scalars. In this approach, the GSF is computed directly from a certain scalar-like self-potential that satisfies the time-domain Teukolsky equation on the Kerr background. The approach is computationally much cheaper than existing time-domain methods, which rely on a direct integration of the linearized Einstein's equations and are impaired by mode instabilities. At the same time, it retains the utility and flexibility of a time-domain treatment, allowing calculations for any type of orbit (including highly eccentric or unbound ones) and the possibility of self-consistently evolving the orbit under the effect of the GSF.  Here we formulate our method, and present a first numerical application, for circular geodesic orbits in Schwarzschild geometry. We discuss further applications.
\end{abstract}

\maketitle

\section{Introduction}

The extreme-mass-ratio regime of the binary black hole inspiral problem is most naturally tackled using the gravitational self-force (GSF) approach, which is based on a systematic expansion of the field equations in the small mass ratio (usually without any weak-field or slow-motion assumptions). At leading order, one has a pointlike particle moving in a geodesic orbit around the large black hole. At 1-GSF order, the particle experiences an effective self-force due to interaction with its own linear gravitational perturbation; nonlinear self-interaction effects are accounted for at subsequent orders. There is now a rigorously established formulation of self-forced motion through 2-GSF order, for general vacuum curved spacetimes (at least for nonspinning particles) \cite{Gralla:2008fg,Gralla:2012db,Poisson:2011nh,Pound:2012nt}. There also exist a number of methods that recast the formal equation of motion into practical regularization schemes in the case of orbits around a Kerr black hole---cardinal among these are mode-sum regularization \cite{Barack:1999wf} and the puncture method \cite{Barack:2007we,Vega:2007mc}; see \cite{Barack:2009ux,Wardell:2015kea} for reviews. Finally, there now exist a variety of computational strategies and working codes that implement the regularization schemes numerically and calculate the GSF for orbits in Schwarzschild or Kerr geometries (see, e.g., \cite{vandeMeent:2016pee} and references therein). Actual calculations have so far been restricted to 1-GSF order, but second-order results are expected soon \cite{Pound:2014xva,Pound:2014koa,Pound:2015wva,Miller:2016hjv}. This computational program is strongly motivated by the prospect of observing extreme-mass-ratio inspirals with planned gravitational-wave observatories in the millihertz band.  

All GSF calculations require information about the local metric perturbation near the particle. Existing schemes differ on the precise type of information required and on how it is obtained in practice. More concretely, schemes  may be classified according to the regularization method they rely on (e.g., mode sum vs.\ puncture), the version of field equations that are being solved (e.g., linearized Einstein's equation vs.\ Teukolsky's equation), or, relatedly, the gauge in which the perturbation is computed. They can also be categorized according to whether the perturbation is solved for in the frequency domain (FD) or in the time domain (TD). In FD methods one solves a single (or a coupled set of) ordinary differential equations (ODEs) for each frequency-harmonic mode of the perturbation. In TD methods one instead directly solves the partial differential equations (PDEs) that govern the time evolution of the perturbation, or of its individual multipole or azimuthal modes.

Two main computational approaches have been responsible for much of the progress in the field so far. The first, and more direct, is based on numerically solving the linearized Einstein's equations in the Lorenz gauge. This can be (and has been) done in conjunction with either mode-sum or puncture regularizations, and is most naturally implemented in the TD. (A FD version of this approach has been developed for calculations in Schwarzschild spacetime \cite{Akcay:2013wfa}, but extending it to Kerr is problematic due to the impossibility of separating the Lorenz-gauge perturbation equations into frequency-harmonic modes in that case.) The Lorenz-gauge/TD approach offers great versatility: codes can be readily implemented for any type of orbit, including highly eccentric or unbound ones, and provide a natural framework for studying the orbital evolution under the GSF effect. But the method has two significant disadvantages. First, it is computationally very expensive: It involves the numerical evolution of a set of 10 coupled PDEs for each multipole or azimuthal mode of the perturbation, over sufficiently long time to ensure that spurious radiation from imperfect initial conditions is sufficiently suppressed. Second, numerical evolutions have been shown \cite{Dolan:2012jg} to be contaminated by certain nonphysical Lorenz-gauge modes that grow linearly in time, the elimination of which  remains an open problem.

The second, less direct computational approach is based on metric reconstruction from curvature scalars. Instead of tackling the complicated set of metric perturbation equations, one works within the elegant framework of the Newman-Penrose formalism, with the numerical task now reduced to solving the Teukolsky equation for either of the two Weyl scalars $\Psi_0$ or $\Psi_4$. Since this equation is fully separable into frequency-harmonic modes, even in the Kerr case, the problem further reduces to solving a set of ODEs in the FD. This approach has been gaining much momentum in the past few years, with the formulation of a metric reconstruction procedure suitable for GSF calculations \cite{Keidl:2006wk,Keidl:2010pm,Merlin:2016boc}, and with the derivation of a 1-GSF equation of motion based on a reconstructed metric \cite{Pound:2013faa}. The Teukolsky/FD approach offers great computational efficiency, and ought to be the method of choice for high-precision calculations. Its main weaknesses are that its efficiency degrades quickly with increasing orbital eccentricity (see \cite{vandeMeent:2016pee} for a detailed discussion of this point) and that its application to orbits that are not strictly periodic is more subtle. In particular, the incorporation of GSF backreaction effects is much less straightforward than in a TD treatment.   

The purpose of this paper is to propose a new computational approach to the GSF, which combines the advantages of the Lorenz-gauge/TD and Teukolsky/FD methods, while avoiding some of their deficiencies. Essentially, our method is a TD application of the metric reconstruction approach. Instead of reconstructing the metric perturbation from a sum over frequency-harmonic modes, as is traditional, we obtain it directly from a certain time-dependent Hertz potential, which, in turn, is obtained by numerically solving the Teukolsky equation in the TD. Our approach thus retains the utility and flexibility of a TD treatment, but offers a much more computationally efficient platform compared to the Lorenz-gauge/TD approach: one solves a single scalarlike equation instead of a coupled set of ten, and the problem of unstable modes is altogether avoided.

TD evolution of the Teukolsky equation has long been used in studies of black hole perturbations, notably by G.\ Khanna and collaborators \cite{LopezAleman:2003ik}. Applications include the study of late-time behavior of vacuum perturbations outside a Kerr black hole \cite{Burko:2013bra}, investigations into the Cauchy horizon singularity inside black holes \cite{Burko:2016uvr}, and the modelling of gravitational radiation from particle orbits \cite{Taracchini:2014zpa}. In these studies, the Teukolsky equation is numerically solved for $\Psi_0$ (or $\Psi_4$) via time evolution in 2+1 dimensions (2+1D), and the relevant physics (e.g., asymptotic behavior at null or timelike infinity) is directly read off that curvature scalar.  In our approach, a time evolution of the Teukolsky equation is used only as a first step in a procedure for reconstructing the local metric perturbation at the particle. Such a direct TD reconstruction has not been attempted so far, to the best of our knowledge.    

Furthermore, in our method we do not solve the Teukolsky equation for the physical Weyl scalar $\Psi_0$ (or $\Psi_4$), as in existing codes. Rather, we solve it for a certain Hertz potential $\Phi$ that is {\it not} the Weyl scalar corresponding to the physical perturbation. The field $\Phi$ satisfies the TD Teukolsky equation in vacuum (with boundary conditions similar to those of the Weyl scalar), but the source term in that equation differs from that of $\Psi_0$ (or $\Psi_4$). A key ingredient of the formulation work to be presented in this paper is a derivation of the point-particle source function for the Hertz potential.\footnote{In related literature, the Hertz potential is usually denoted by $\Psi$, with a lowercase $\psi$ denoting the Weyl scalars: $\psi_0$ and $\psi_4$. Here, to avoid potential confusion, we use $\Phi$ for the Hertz potential; lowercase $\phi$ and $\psi_0/\psi_4$ will be reserved for the 1+1D projections of $\Phi$ and $\Psi_0/\Psi_4$, respectively.}

Our method is designed with a 1+1D implementation in mind, that is a numerical evolution (either Cauchy or characteristic) on a 2D grid with one temporal dimension and one spatial dimension. In the case of a  Schwarzschild background, the Teukolsky equation naturally separates into decoupled 1+1D evolution equations for each multipole (spin-weighted spherical harmonic) mode of $\Phi$. Not so in the Kerr case, where different multipoles ($\ell$ modes) remain coupled. Nonetheless, a 1+1D implementation is still a viable route even in the Kerr case, as demonstrated in Ref.\ \cite{Dolan:2012yt}, at least in situations where mode coupling is relatively weak.  In a 1+1D treatment, the particle's worldline splits the numerical grid into two disjoint domains. The formulation of a source term for $\Phi$ then translates into the prescription of {\em junction} conditions that the field $\Phi$ must satisfy on the interface between the two domains---specifically, the jumps in the value of $\Phi$ and a sufficient number of its derivatives across the particle's worldline. The bulk of our formulation work will be devoted to deriving these junction conditions.    

To recap, our goal here is to formulate a TD evolution problem for a certain Hertz potential $\Phi$, from which the physical metric perturbation and the GSF can be derived directly by taking derivatives. We will formulate our method for arbitrary particle orbits in Kerr spacetime, and then, as a proof of principle, we shall present a full numerical implementation for the case of circular geodesic orbits in Schwarzschild spacetime.  We leave the full numerical implementation of our method in Kerr to future work.

The structure of the paper is as follows. In Sec.\ \ref{Sec:GSF} we review the derivation of the GSF from a reconstructed metric, as proposed by Pound {\it et al.}\ in \cite{Pound:2013faa}. We also review the reconstruction procedure starting from a Hertz potential $\Phi$. This will provide the necessary formal background for our work. Section \ref{Sec:scheme} describes our new method and formulates a practical TD evolution scheme for $\Phi$. This is done for generic orbits in Kerr spacetime. Section \ref{Sec:example} specializes to circular geodesic orbits in Schwarzschild spacetime, giving all necessary implementation formulas in explicit form for that case. In Sec.\ \ref{Sec:numerics} we then present an illustrative numerical implementation of the method, for circular geodesic orbits in Schwarzschild.  We conclude in Sec.\ \ref{Sec:summary} by laying out of a program for the full numerical implementation of our method in Kerr.  

Throughout this work we use standard geometrized units, in which $G=1=c$. We adopt the metric signature $({-}{+}{+}{+})$. Greek letters are used for spacetime indices, and a comma denotes a partial derivative, as in $A_{,\alpha}:= \prt A/\prt x^\alpha$. Boldface roman indices, as in $e^\alpha_{\bf a}$, run over $1,\ldots, 4$ and identify tetrad legs.   Complex conjugation is denoted by an overbar, as in $\bar{m}^{\alpha}$. Parenthetical indices are symmetrized, as in $A_{(\alpha\beta)}=(A_{\alpha\beta}+A_{\beta\alpha})/2$.
We will consider a Kerr background with metric $g^K_{\alpha\beta}$, mass parameter $M$ and angular-momentum parameter $a$. We will usually adopt standard Boyer-Lindquist coordinates $(t,r,\theta,\varphi)$. Our sign conventions for the Weyl scalars, the spin coefficients of the Newman-Penrose formalism, and the Hertz potential, are consistent with those of Ref.\ \cite{Merlin:2016boc} (as summarized in Appendix A therein).

\section{Self-force from a reconstructed metric: a review} \label{Sec:GSF}

This section reviews the relevant background, while setting up notation and conventions.  We first describe the (TD version of the) standard procedure for reconstruction of vacuum metric perturbations in Kerr geometry, then review the case where the perturbation is sourced by an orbiting point particle, and finally  summarize the method by Pound {\it et al.}.\ \cite{Pound:2013faa} for computing the GSF experienced by the particle, using the reconstructed metric as input. 

\subsection{Metric reconstruction in vacuum}

A reconstruction procedure for vacuum perturbations was developed long ago by Chrzanowski \citep{Chrzanowski:1975wv} and Cohen and Kegeles \citep{Kegeles:1979an}, with later contributions from Wald \citep{Waldrec}, Stewart \citep{Stewart:1978tm}, Lousto and Whiting \citep{Lousto:2002em} and others. As is common, we shall refer to it here as the CCK procedure. Considering a vacuum perturbation $h_{\alpha\beta}$ of a Kerr black hole geometry, with corresponding Weyl scalars $\Psi_0$ and $\Psi_4$, the CCK method prescribes the reconstruction of $h_{\alpha\beta}$ from either $\Psi_0$ or $\Psi_4$. More precisely, the reconstruction procedure returns a perturbation $h^{\rm rec}_{\alpha\beta}$ that is equal to $h_{\alpha\beta}$ up to (i) some gauge perturbation $h^{\rm gauge}_{\alpha\beta}$, and (ii) a ``completion'' piece $h^{\rm comp}_{\alpha\beta}$ representing a four-parameter family of simple, stationary and axisymmetric vacuum solutions: mass and angular momentum perturbations of Kerr, and perturbations away from Kerr into Kerr-NUT or C-metric geometries \cite{waldtheo}. Thus, the original perturbation is given by
\begin{equation}\label{completed}
h_{\alpha\beta}=h^{\rm rec}_{\alpha\beta}+h^{\rm comp}_{\alpha\beta}+h^{\rm gauge}_{\alpha\beta},
\end{equation}
where $h^{\rm comp}_{\alpha\beta}$ and $h^{\rm gauge}_{\alpha\beta}$ are not fixed within the CCK procedure, and must be determined separately. 

There are two variants of the reconstruction procedure, returning $h^{\rm rec}_{\alpha\beta}$ in two different gauges, known as the ``ingoing'' and ``outgoing'' traceless radiation gauges (IRG and ORG, respectively). The corresponding gauge conditions are
\begin{equation}
l^\alpha h^{\rm rec}_{\alpha\beta}=0\ \ \text{(IRG)}, \quad\quad
n^\alpha h^{\rm rec}_{\alpha\beta}=0\ \ \text{(ORG)},
\end{equation}
along with the trace-free condition
\begin{equation}
g_K^{\alpha\beta}h^{\rm rec}_{\alpha\beta}=0,
\end{equation}
where $g_K^{\alpha\beta}$ is the inverse of the background Kerr metric $g^K_{\alpha\beta}$.
Here we have introduced Kinnersley's null tetrad $e^{\alpha}_{\bf a}$ (${\bf a}=1,2,3,4$), whose legs are given, in Boyer-Lindquist coordinates, by 
\begin{subequations} \label{eq:kerrtetrad}
\begin{align}
e_{\bf 1}^{\alpha} =: l^\alpha &=\frac{1}{\Delta}\left(r^2+a^2,\Delta,0,a\right),\\
e_{\bf 2}^{\alpha} =: n^\alpha &=\frac{1}{2\Sigma}\left(r^2+a^2,-\Delta,0,a \right),\\
e_{\bf 3}^{\alpha} =: m^\alpha &=-\frac{\bar\varrho}{\sqrt{2}}\left(ia\sin\theta,0,1,\frac{i}{\sin\theta}\right), \\
e_{\bf 4}^{\alpha} =: \bar m^\alpha &=\frac{\varrho}{\sqrt{2}}\left(ia\sin\theta,0,-1,\frac{i}{\sin\theta}\right),
\end{align}
\end{subequations}
where
\begin{eqnarray}
\Delta&:=&r^2-2Mr+a^2, \nonumber\\
\Sigma&:=&r^2+a^2\cos^2\theta, \nonumber\\
\varrho&:=& -1/(r-ia\cos\theta).
\end{eqnarray}
The legs $e^{\alpha}_{\bf n}$ are all null and mutually orthogonal, except $l^\alpha n_\alpha=-1$ and  $m^\alpha\bar{m}_\alpha=1$. 

The CCK procedure starts with the derivation of a suitable {\it Hertz potential} $\Phi$ (one such potential for ORG and another for IRG---call them $\Phi^{\rm ORG}$ and $\Phi^{\rm IRG}$, respectively). This Hertz potential is required to satisfy two differential equations. The first is the master (source-free) Teukolsky equation,
\begin{align}
\label{eq:kerrteuk} 
&\left(\frac{(r^2+a^2)^2}{\Delta} -a^2\sin^2\theta\right)\frac{\partial^2\Phi}{\partial t^2}
+\frac{4Mar}{\Delta}\frac{\partial^2\Phi}{\partial t\partial\varphi}
\nonumber\\
&+\left(\frac{a^2}{\Delta}-\frac{1}{\sin^2\theta}\right)\frac{\partial^2\Phi}{\partial\varphi^2}
-\Delta^{-s}\frac{\partial}{\partial r}\left(\Delta^{s+1}\frac{\partial\Phi}{\partial r}\right)
\nonumber\\
&-\frac{1}{\sin\theta}\frac{\partial}{\partial\theta}\left(\sin\theta \frac{\partial\Phi}{\partial\theta}\right) +(s^2\cot^2\theta -s)\Phi
\nonumber\\ 
&-2s\left(\frac{M(r^2-a^2)}{\Delta}-r-ia\cos\theta\right)\frac{\partial\Phi}{\partial t} 
\nonumber \\
&\hspace{5.0em}-2s\left(\frac{a(r-M)}{\Delta}+\frac{i\cos\theta}{\sin^2\theta}\right)\frac{\partial\Phi}{\partial\varphi}=0 ,
\end{align}
with $s=+2$ for $\Phi^{\rm ORG}$ and $s=-2$ for $\Phi^{\rm IRG}$. The second equation links $\Phi$ to one of the given Weyl-scalar perturbations $\Psi_0$ or $\Psi_4$:
\begin{eqnarray}
{\boldsymbol D}_l^4\bar\Phi^{\rm IRG} &=& 2\Psi_0, 
\label{invertion_rad_IRG}
\\
\Delta^{2}\tilde{\boldsymbol D}_n^4\Delta^2\bar\Phi^{\rm ORG} &=& 32\varrho^{-4}\Psi_4 ,
\label{invertion_rad_ORG}
\end{eqnarray}
where 
\begin{eqnarray}
\boldsymbol{D}_l&:=&l^{\alpha}\prt_{\alpha}=\frac{r^2+a^2}{\Delta}\partial_t+\partial_r+\frac{a}{\Delta}\partial_{\varphi},
\nonumber\\ 
\tilde{\boldsymbol D}_n&:=&-\frac{2\Sigma}{\Delta}n^{\alpha}\prt_{\alpha}=-\frac{r^2+a^2}{\Delta}\partial_t+\partial_r-\frac{a}{\Delta}\partial_{\varphi},
\end{eqnarray}
and $\boldsymbol{D}_l^4:=\boldsymbol{D}_l\boldsymbol{D}_l\boldsymbol{D}_l\boldsymbol{D}_l$, etc.
Alternatively, instead of (\ref{invertion_rad_IRG}) or (\ref{invertion_rad_ORG}), one can require
\begin{eqnarray}
\tilde{\cal L}_{-1}\tilde{\cal L}_0\tilde{\cal L}_{1}\tilde{\cal L}_{2}\bar\Phi^{\rm IRG}-12M\Phi^{\rm IRG}_{,t} &=& 8\varrho^{-4}\Psi_4, 
\label{invertion_ang_IRG}
\\
{\cal L}_{1}{\cal L}_0{\cal L}_{-1}{\cal L}_{-2}\bar\Phi^{\rm ORG}+12M\Phi^{\rm ORG}_{,t} &=& 8\Psi_0, 
\label{invertion_ang_ORG}
\end{eqnarray}
where
\begin{eqnarray}
{\cal L}_s&:=& -\left(\partial_\theta-s\cot\theta+i\csc\theta\partial_\varphi\right)-ia\sin\theta\partial_t,
\nonumber\\ 
\tilde{\cal L}_s&:=&-\left(\partial_\theta+s\cot\theta-i\csc\theta\partial_\varphi\right)+ia\sin\theta\partial_t.
\end{eqnarray}
In can be shown \cite{Ori:2002uv} that the combination of (\ref{eq:kerrteuk}) and (\ref{invertion_rad_IRG}) [or (\ref{eq:kerrteuk}) and (\ref{invertion_ang_IRG})] determines $\Phi^{\rm IRG}$ uniquely, and similarly the combination of (\ref{eq:kerrteuk}) and (\ref{invertion_rad_ORG}) [or (\ref{eq:kerrteuk}) and (\ref{invertion_ang_ORG})] determines $\Phi^{\rm ORG}$ uniquely.

Given $\Phi$, the vacuum perturbation $h^{\rm rec}_{\alpha\beta}$ is obtained via
\begin{equation}\label{reconstruction}
h^{\rm rec}_{\alpha\beta}={\rm Re}\left( e_{{\bf a}(\alpha}e_{{\bf b}\beta)}{\cal D}^{\bf ab}\Phi\right),
\end{equation}
where the symmetrization is over the tensorial indices $\alpha\beta$, and ${\cal D}^{\bf ab}$ are certain second-order differential operators. Explicitly, the nonvanishing operators ${\cal D}^{\bf ab}$ are given by 
\begin{eqnarray}
{\cal D}^{\bf 11}&=&-2({\boldsymbol D}_m+\bar\alpha+3\beta-\tau)({\boldsymbol D}_m+4\beta+3\tau),
\nonumber\\
{\cal D}^{\bf 33}&=& -2({\boldsymbol D}_l-\varrho)({\boldsymbol D}_l+3\varrho),
\nonumber\\
{\cal D}^{\bf 13}&=&{\cal D}^{\bf 31}=({\boldsymbol D}_m-2\bar\alpha+2\beta-\tau)({\boldsymbol D}_l+3\varrho)
\nonumber\\&& \hspace{12mm}
+ ({\boldsymbol D}_l+\bar\varrho-\varrho)({\boldsymbol D}_m+4\beta+3\tau)
\end{eqnarray}
for the IRG, and by 
\begin{eqnarray}
{\cal D}^{\bf 22}&=& -2\varrho^{-4}({\boldsymbol D}_{\bar m}+2\alpha+4\bar\beta-\tau)({\boldsymbol D}_{\bar m}-3\alpha+\bar\beta),
\nonumber\\
{\cal D}^{\bf 44}&=& -2\varrho^{-4}({\boldsymbol D}_n+5\mu-3\gamma+\bar\gamma)({\boldsymbol D}_n+\mu-4\gamma),
\nonumber\\
{\cal D}^{\bf 24}&=&{\cal D}^{\bf 42}=\varrho^{-4}\left[({\boldsymbol D}_{\bar m}+2\alpha+4\bar\beta+\bar\tau)({\boldsymbol D}_n+\mu-4\gamma)\right.
\nonumber\\&& 
+ \left. ({\boldsymbol D}_n+5\mu-\bar\mu-3\gamma-\bar\gamma)({\boldsymbol D}_{\bar m}-3\alpha+\bar\beta)\right]
\end{eqnarray}
for the ORG. Here we have introduced
\begin{eqnarray}
\boldsymbol{D}_n&:=&n^{\alpha}\prt_{\alpha}=\frac{1}{2\Sigma}\left[(r^2+a^2)\partial_t-\Delta\partial_r+a\partial_{\varphi}\right],
\nonumber\\ 
{\boldsymbol D}_m&:=&m^{\alpha}\prt_{\alpha}=-\frac{\bar\varrho}{\sqrt{2}}\left(ia\sin\theta\partial_t+\partial_\theta+i\csc\theta\partial_{\varphi}\right),
\nonumber\\ 
{\boldsymbol D}_{\bar m}&:=&\bar m^{\alpha}\prt_{\alpha}=
\frac{\varrho}{\sqrt{2}}\left(ia\sin\theta\partial_t-\partial_\theta+i\csc\theta\partial_{\varphi}\right),
\end{eqnarray}
and
\begin{eqnarray}
\beta&:=&-\frac{\bar{\varrho}\cot\theta,}{2\sqrt{2}},
\quad\quad
\alpha:=\frac{ia\varrho^2\sin\theta}{\sqrt{2}}-\bar\beta,
\nonumber\\
\mu&:=&\frac{\varrho\Delta}{2\Sigma}, \quad\quad
\gamma:= \mu+\frac{r-M}{2\Sigma},
\nonumber\\
\tau&:=&-\frac{ia\sin\theta}{\sqrt{2}\Sigma}.
\end{eqnarray}

\subsection{Metric reconstruction for a particle source}

The above reconstruction procedure is guaranteed to return a valid vacuum solution $h_{\alpha\beta}^{\rm rec}$ whenever the Weyl scalar one starts with ($\Psi_4$ or $\Psi_0$) is a solution to the source-free Teukolsky equation. But when matter sources are present, the procedure can fail to return a valid solution even at vacuum points away from any sources. This is true even in the simplest example of a static particle in flat space, in either the IRG or the ORG. As shown first in \cite{Barack:2001ph}, the reconstructed perturbation develops a stringlike singularity, which emanates from the particle in the radial null direction, either outward or inward, or in both directions (depending on the specific choice of gauge within each of the IRG or ORG classes). Reference \cite{Pound:2013faa} introduced a categorization of reconstructed perturbations based on the form of singularity---the two ``half-string'' classes and the ``full string'' class---and showed that any reconstructed metric belongs to one of the classes (assuming continuity away from the string). A ``no-string'' gauge may be formed by joining together the two ``regular sides'' of two opposite half-string perturbations along a closed surface $\cal S$ through the particle. Such a construction was first introduced by Friedman and collaborators in \cite{Keidl:2006wk,Keidl:2010pm}. The no-string perturbation is free from stringlike singularities, but has a gauge discontinuity (and also delta-function distributions \cite{Pound:2013faa}) on the interface $\cal S$.

Specializing to a particle in a bound orbit around a Kerr black hole, we let the particle's worldline be represented (in Boyer-Lindquist coordinates) by $x_{\rm p}=\{t,r_{\rm p}(t),\theta_{\rm p}(t),\varphi_{\rm p}(t)\}$, where the radius $r_{\rm p}(t)$ is compactly supported outside the Black hole's event horizon. It is convenient to choose the interface $\cal S$ to be the surface $r=r_{\rm p}(t)$, and we denote the interior of $\cal S$ [i.e., $r<r_{\rm p}(t)$] by ${\cal S}^-$ and its exterior [i.e., $r>r_{\rm p}(t)$] by ${\cal S}^+$. Then a half-string reconstructed metric has a string singularity in either ${\cal S}^-$ or ${\cal S}^+$ (and is smooth elsewhere off the particle), and a full-string metric has a string singularity in both. A no-string reconstructed metric is smooth anywhere in both vacuum domains ${\cal S}^-$ and ${\cal S}^+$ but has a discontinuity across the interface $\cal S$ between them.

Our time-domain metric reconstruction prescription in this work will yield the no-string vacuum perturbations $h_{\alpha\beta}^{\rm rec\pm}$, where henceforth in this paper $\pm$ denotes values in the corresponding vacuum domains ${\cal S}^{\pm}$. We denote the corresponding total, completed perturbations in ${\cal S}^{\pm}$ by
\begin{equation}\label{completed_pm}
h_{\alpha\beta}^{\pm}=h^{\rm rec\pm}_{\alpha\beta}+h^{\rm comp\pm}_{\alpha\beta}+h^{\rm gauge\pm}_{\alpha\beta},
\end{equation}
 where $h^{\rm comp\pm}_{\alpha\beta}$ and $h^{\rm gauge\pm}_{\alpha\beta}$ are the contributions from the ``completion'' and ``gauge'' perturbations in the corresponding domains [recall Eq.\ (\ref{completed})]. The completion piece $h^{\rm comp\pm}_{\alpha\beta}$ has been derived recently in Refs.\ \cite{Merlin:2016boc,vandeMeent:2017fqk} (for arbitrary bound geodesic motion in Kerr spacetime), and is given there in explicit analytic form.
A gauge adjustment $h^{\rm gauge\pm}_{\alpha\beta}$ may need to be included in certain applications (see Refs.\ \cite{Shah:2015nva} or \cite{vandeMeent:2016hel} for examples), but in this work we will set $h^{\rm gauge\pm}_{\alpha\beta}\equiv 0$ for simplicity. Thus, our attention in the rest of this paper will be focused on the construction of the perturbation $h^{\rm rec\pm}_{\alpha\beta}$.

\subsection{Self-force from a reconstructed metric}\label{subsec:mode-sum}

In Ref.\ \cite{Pound:2013faa}, Pound \textit{et al.}~obtained a formulation of the GSF starting from a reconstructed metric (in either the IRG or the ORG), complete with practical mode-sum formulas. Two different schemes were described, one based on (either of the two) half-string gauges, and another based on the no-string gauge. The prescription for calculating the GSF in the no-string scheme is as follows. First, given the reconstructed (and completed) perturbations $h_{\alpha\beta}^{\pm}$, introduce the one-sided ``force'' fields 
\qwe \label{Fpm}
F_{\alpha}^{\pm}:=-\frac{1}{2}\mu(\delta_{\alpha}^{\beta}+u_{\alpha}u^{\beta})(2\nabla_{\nu} h_{\beta\mu}^{\pm}-\nabla_{\beta} h_{\mu\nu}^{\pm})u^{\mu}u^{\nu},
\ewq
defined on ${\cal S}^{\pm}$, respectively,
where $\mu$ is the mass of the particle, $\nabla_{\alpha}$ is a covariant derivative compatible with the background (Kerr) geometry, and $u^{\alpha}$ is any smooth extension of the particle's four-velocity to form a vector field in spacetime. Next, expand each coordinate component of the fields $F_{\alpha}^{\pm}$ in spherical harmonics $Y_{\ell m}(\theta,\varphi)$ on spheres of constant $r,t$, and for each multipole contribution (summed over azimuthal number $m$) take the one-sided radial limits to the particle:   
\qwe \label{Flpm}
F_{\alpha}^{\ell\pm}:=\lim_{r\to r_{\rm p}^{\pm}}\sum_{m=-\ell}^\ell Y_{\ell m}(\theta_{\rm p},\varphi_{\rm p})\int F_{\alpha}^{\pm} \bar Y_{\ell m}(\theta,\varphi)d\Omega , 
\ewq 
where $d\Omega:=\sin\theta d\theta d\varphi$. These one-sided $\ell$-mode force contributions are each finite (bounded), and generally grow in amplitude as $\sim \ell$ for large $\ell$. According to \cite{Pound:2013faa}, given $F_{\alpha}^{\ell\pm}$, the physical GSF is calculated via  
\qwe\label{mode sum}
F_{\alpha}=\sum_{\ell=0}^{\infty} \left[\frac{1}{2}(F^{\ell+}_{\alpha}+F^{\ell-}_{\alpha})-B_{\alpha}\right],
\ewq
where $B_{\alpha}$ is the standard, Lorenz-gauge regularization parameter (which depends on the details of the orbit, but not on $\ell$). An analytical expression for $B_{\alpha}$, for generic orbits in Kerr, was first derived in \cite{Barack:2002mh} and it is given explicitly in \cite{Barack:2009ux}. 

The mode-sum formula (\ref{mode sum}) can be implemented in either the IRG or the ORG. While $F^{\ell\pm}\sim \ell$ at large $\ell$, the average $\frac{1}{2}(F^{\ell+}_{\alpha}+F^{\ell-}_{\alpha})$ approaches a constant ($\ell$-independent) value, and the entire summand in (\ref{mode sum}) is guaranteed to fall off at least as $\sim \ell^{-2}$. Hence, the mode sum converges at least as $\sim 1/\ell$.

In conclusion, we see that knowledge of the ``no-string'' perturbations $h^{\rm rec\pm}_{\alpha\beta}$ (and their derivatives at the particle) provides sufficient input for calculating the physical GSF. Since $h^{\rm rec\pm}_{\alpha\beta}$ are each a vacuum solution in its corresponding vacuum domain (${\cal S}^{\pm}$), they can be reconstructed from suitable Hertz potentials $\Phi^{\pm}$ by applying the vacuum reconstruction formula (\ref{reconstruction}) in each of the domains:
\begin{equation}\label{reconstruction_pm}
h^{\rm rec\pm}_{\alpha\beta}={\rm Re}\left( e_{{\bf a}(\alpha}e_{{\bf b}\beta)}{\cal D}^{\bf ab}\Phi^{\pm}\right).
\end{equation}
In the next section we will lay out our proposed method for deriving the appropriate potentials $\Phi^{\pm}$, via a direct time-domain evolution of the Teukolsky equation. Once $\Phi^{\pm}$ are at hand, the GSF is obtained (in a completed no-string gauge) by substituting for $\Phi^{\pm}$ in Eq.\ (\ref{reconstruction_pm}), then consequently for $h^{\rm rec\pm}_{\alpha\beta}$ in Eq.\ (\ref{completed_pm}),
for $h^{\pm}_{\alpha\beta}$ in Eq.\ (\ref{Fpm}), for $F_{\alpha}^{\pm}$ in Eq.\ (\ref{Flpm}), and finally for $F_{\alpha}^{\ell\pm}$ in the mode-sum formula (\ref{mode sum}).

\section{1+1D evolution scheme for the Hertz potentials $\Phi^{\pm}$}\label{Sec:scheme}

Our goal here is to formulate a practical evolution scheme for $\Phi^{\pm}$, suitable for numerical integration in 1+1D. Such a formulation requires three components. First, starting from the master Teukolsky equation (\ref{eq:kerrteuk}), we need to introduce a suitable decomposition of $\Phi^{\pm}$ into multipole modes and obtain a time-evolution equation for the ``time-radial'' piece of each of the modes. Second, we need to derive the physical boundary conditions for the fields $\Phi^{\pm}$ and for their multipole modes.
Third, we need to formulate junction conditions for the multipole modes on the surface $\cal S$.  These three components of the problem are dealt with, each in turn, in the next three subsections.  

We aim here to give a full formulation for both $\Phi^{\rm IRG}$ and $\Phi^{\rm ORG}$. Since the former is a solution of the master equation  (\ref{eq:kerrteuk}) with $s=-2$, and the latter is a solution of that equation with $s=+2$, we shall use the notation 
\begin{equation}
\Phi_{-2}:=\Phi^{\rm IRG}, \quad\quad
\Phi_{+2}:=\Phi^{\rm ORG},
\end{equation}
which will allow us to unify the treatment of both gauges. Correspondingly, the vacuum potentials in ${\cal S}^{\pm}$ will be denoted by $\Phi_s^\pm$, with $s=-2$ for IRG and $s=+2$ for ORG. 

\subsection{The Teukolsky equation in a 1+1D form }

As far as we know, in general, the master Teukolsky equation (\ref{eq:kerrteuk}) cannot be separated in the time domain, i.e.\ without first decomposing $\Phi_s$ into frequency modes $\sim e^{i\omega t}$; only the azimuthal dependence can be separated, using $\Phi_s=\sum_m\Phi_{sm}e^{im\varphi}$. A full separation of the angular dependence becomes possible in the special case of $a=0$ (Schwarzschild background), or for $t$-independent perturbations. In both special cases, separation is achieved using a basis of {\em spin-weighted spherical harmonic} functions ${}_s\!Y_{\ell m}(\theta,\varphi)$ \cite{Goldberg1967}. For spins $s=\pm 2$, relevant here, these functions are derived from the standard spherical harmonics $Y_{\ell m}(\theta,\varphi)$ via 
\begin{align}
{}_{\pm 2}\!Y_{\ell m}=\sqrt{\frac{(\ell-2)!}{(\ell+2)!}}\left[\frac{\partial^2 Y_{\ell m}}{\partial\theta^2}
-\left(\frac{\cos\theta\pm 2m}{\sin\theta}\right)\frac{\partial Y_{\ell m}}{\partial\theta}\right.
\nonumber\\
+\left.\left(\frac{m^2\pm 2m\cos\theta}{\sin^2\theta}\right)Y_{\ell m}\right],
\end{align}
and they satisfy the differential equation
\begin{align}\label{YslmEq}
\frac{1}{\sin\theta}\frac{\partial}{\partial\theta}\left(\sin\theta\frac{\partial {}_s\!Y_{\ell m}}{\partial\theta}\right) 
+\left(-\frac{m^2+2ms\cos\theta}{\sin^2\theta}\right.
\nonumber\\
 -s^2\cot^2\theta+s+(\ell-s)(\ell+s+1)\Big){}_s\!Y_{\ell m}=0 .
\end{align}
Since the spherical harmonics satisfy $\bar Y_{\ell m}\equiv(-1)^m Y_{\ell,-m}$ (where the sign factor is conventional), it is easy to see that we have the symmetry relation
\begin{equation}\label{symmetry}
{}_{\pm 2}\!\bar Y_{\ell m}\equiv(-1)^m {}_{\mp 2}\!Y_{\ell,-m}.
\end{equation}

Our strategy here will be to expand $\Phi_s^{\pm}$ in ${}_s\!Y_{\ell m}(\theta,\varphi)$ even in the Kerr case. The resulting field equations for the time-radial modes will exhibit coupling between modes of different $\ell$ (though modes of different $m$ will remain decoupled). We will then formulate our evolution problem for that mode-coupled set. 

We thus expand the fields $\Phi_s^{\pm}$ in the form
\begin{equation} \label{expansion}
\Phi_s^{\pm}=
(r\Delta^{s})^{-1}
       \sum_{\ell=2}^{\infty}\sum_{m=-\ell}^{\ell}
       \phi^{\pm}_{s\ell m}(t,r) {}_s\!Y_{\ell m}(\theta,\tilde\varphi).
\end{equation}
Here, the radial factor $(r\Delta^{s})^{-1}$ is introduced to regulate the behavior of 
$\phi^{\pm}_{s\ell m}$ at infinity and on the event horizon; its particular form will be explained in Sec.\ \ref{subsec:BC} below. The azimuthal coordinate $\tilde\varphi$ is a horizon-regularized version of the Boyer-Lindquist $\varphi$, defined through
\begin{equation}
\tilde\varphi = \varphi +\frac{a}{r_+-r_-}\ln\left(\frac{r-r_+}{r-r_-}\right),
\end{equation}
where
\begin{equation}
r_\pm= M\pm (M^2-a^2)^{1/2};
\end{equation}
it satisfies $\partial\tilde\varphi/\partial r=a/\Delta$. Had we instead used the standard $\varphi$ in Eq.\ (\ref{expansion}), the field $\phi^{-}_{s\ell m}$ would exhibit irregular oscillations $\sim e^{\pm im\Omega_{\rm H} r_*}$ near the event horizon \cite{Teukolsky:1972my}. Here $\Omega_{\rm H}=a/(2Mr_+)$ is the horizon's angular velocity, and $r_*$ is the standard ``tortoise'' coordinate, satisfying 
\begin{equation}
\frac{dr_*}{dr}=\frac{r^2+a^2}{\Delta}.
\end{equation}
Note that $\tilde\varphi=\varphi +O(a/r)$, and hence $\tilde\varphi$ approaches the standard $\varphi$ at large $r$. 

Inserting the expansion (\ref{expansion}) into the master equation (\ref{eq:kerrteuk}), and using Eq.\ (\ref{YslmEq}), we obtain
\begin{align} \label{NoSeparation}
\sum_{\ell m}{}_s\!Y_{\ell m}(\theta,\tilde\varphi)\left[\tilde{D}\phi^{\pm}_{s\ell m}
- a^2 \sin^2\theta\,(\phi_{s\ell m}^{\pm})_{,tt}\right.
\nonumber\\
+ \left.2ias\cos\theta\,(\phi^{\pm}_{s\ell m})_{,t}\right]=0,
\end{align}
where $\tilde{D}$ is a certain partial differential operator independent of $\theta,\varphi$. Note how the two terms $\propto \sin^2\theta$ and $\propto \cos\theta$ prevent a full separation of variables when $a\ne 0$ and the field is $t$ dependent. Following the treatment (and notation) of Ref.\ \cite{Barack:1999st}, we proceed by re-expanding the angular functions $ {}_s\!Y_{\ell m}\cos\theta$ and ${}_s\!Y_{\ell m}\sin^2\theta $ in spin-weighted harmonics:
\begin{align} \label{cosYexpansion}
{}_s\!Y_{\ell} \cos\theta=
c_{-}^{\ell+1} {}_s\!Y_{\ell+1} + c_0^\ell\, {}_s\!Y_{\ell} + c_{+}^{\ell-1} {}_s\!Y_{\ell-1},
\end{align}
\begin{align} \label{sin2Yexpansion}
 {}_s\!Y_{\ell} \sin^2\theta=
  C_{--}^{\ell+2} {}_s\!Y_{\ell+2}
+ C_{-}^{\ell+1}     {}_s\!Y_{\ell+1}
+ C_0^\ell         {}_s\!Y_{\ell}
\nonumber\\
+ C_{+}^{\ell-1}    {}_s\!Y_{\ell-1}
+ C_{++}^{\ell-2}  {}_s\!Y_{\ell-2}.
\end{align}
Here we have dropped the index $m$ for clarity, and the coefficients are
\begin{equation} \label{cosYcoefficients}
\begin{array}{l}
c_-^\ell = \left[\frac{\left(\ell^2-s^2\right)\left(\ell^2-m^2\right)}
                       {\ell^2 (2\ell-1)(2\ell+1)}\right]^{1/2}, \\
c_0^\ell = -\frac{ms}{\ell(\ell+1)}, \\
c_+^\ell = c_-^{\ell+1} ,
\end{array}
\end{equation}
and
\begin{equation} \label{sin2Ycoefficients}
\begin{array}{lcl}
    C_{++}^\ell &=&  -c_+^{\ell+1} c_+^\ell,                      \\
    C_{+}^\ell    &=&  -c_+^\ell (c_0^{\ell+1} + c_0^\ell),            \\
    C_0^\ell    &=& 1-(c_-^\ell)^2-(c_+^\ell)^2-(c_0^\ell)^2,        \\
    C_{-}^\ell    &=&  -c_-^\ell (c_0^\ell + c_0^{\ell-1}),            \\
    C_{--}^\ell &=&  -c_-^{\ell-1} c_-^\ell .
\end{array}
\end{equation}
Substituting back into (\ref{NoSeparation}) and using the orthogonality property of the functions ${}_s\!Y_{lm}$, we finally separate out the angular dependence, arriving at a mode-coupled set of equations for the time-radial part. For each $s,\ell,m$ (with $s=\pm 2$, $\ell\geq 2$ and $|m|\leq \ell$), it has the form 
\begin{equation} \label{psicoupled}
\hat{D}\phi_{sm}^{\ell}  + {\cal I}(\phi_{sm}^{\ell\pm1},\phi_{sm}^{\ell\pm2})=0,
\end{equation}
where, to avoid confusion, we have omitted the label $\pm$ associated with the domains ${\cal S}^{\pm}$, and used superscript for the multipole label $\ell$. In this equation, $\hat{D}$ is yet another time-radial differential operator (independent of $\theta,\varphi$), and $\cal I$ is a functional describing {\em coupling} between each of the $\ell$-modes and their nearest and next-to-nearest neighbours:
\begin{align} \label{calI}
{\cal I}=
-a^2\left(C_{++}^\ell\phi^{\ell+2}_{sm} 
+C_+^\ell\phi^{\ell+1}_{sm}
+C_-^\ell\phi^{\ell-1}_{sm}
+C_{--}^\ell\phi^{\ell-2}_{sm}\right)_{,tt}
\nonumber\\
+2ias\left(c_+^\ell\phi^{\ell+1}_{sm} + c_-^\ell\phi^{\ell-1}_{sm}\right)_{,t}.
\end{align}
The coupling disappears when $a=0$ or the perturbation is $t$ independent. 

To write Eq.\ (\ref{psicoupled}) explicitly in a convenient form, we introduce the advanced and retarded time coordinates,
\begin{eqnarray} \label{EF}
v:= t+r_* \quad\text{and}\quad u:= t-r_* ,
\end{eqnarray}
respectively (which reduce to the standard double-null Eddington-Finkelstein coordinates in the Schwarzschild case).
Then the 1+1D modal Teukolsky equation (\ref{psicoupled}) takes the explicit form
\begin{align} \label{Teukolsky1+1}
\phi^\ell_{,uv} + U(r)\phi^\ell_{,u} + V(r)\phi^\ell_{,v}  + W(r)\phi^\ell 
\hspace{15mm}
\nonumber\\ 
+K(r)\left[-a^2 C_0^\ell \,\phi^\ell_{,tt}
+{\cal I}(\phi^{\ell\pm1},\phi^{\ell\pm2})\right]=0,
\end{align}
where $\partial_u$, $\partial_v$ and $\partial_t$ are taken with fixed $v$, $u$ and $r$, respectively, and we have dropped the indices $s,m$ for improved readability. The radial functions in this equation read
\begin{equation} \label{K}
K(r)=\frac{\Delta}{4(r^2+a^2)^2},
\end{equation}
\begin{align} \label{U}
U(r)=2K(r)\left[2sM+ia(sc_0^\ell+m)-a^2/r\right.
\nonumber\\
\left. +4Mr(sM-sr+iam)/\Delta \right],
\end{align}
\begin{equation} \label{V}
V(r)=2K(r)\left[2sr+ia(sc_0^\ell-m)+a^2/r\right],
\end{equation}
\begin{align} \label{W}
W(r)=K(r)\left[
(\ell-s)(\ell+s+1)+2(s+1)M/r\right.
\nonumber\\
\left.+2iam/r-2a^2/r^2
\right].
\end{align}

It is important to reiterate that, even in the Kerr case, the 1+1D Teukolsky equation (\ref{Teukolsky1+1}) exhibits only a {\em finite} coupling between spherical-harmonic modes: each $\ell$ mode couples only to its nearest and next-to-nearest neighbors. This is a remarkable property that can bring much simplification in practice (see below). We also note that the coupling terms contained in $\cal I$ in Eq.\ (\ref{Teukolsky1+1}) are expected to be subdominant (compared to, e.g., the term $\propto W$ in that equation) in the problem of interest to us here: these terms are relatively suppressed by factors of orders $(a\omega)^2$ or $a\omega$, where $\omega$  is a characteristic frequency of the perturbation, which, in most relevant scenarios (and for relevant values of $m$) is considerably smaller than $1/M$, giving suppression factors considerably smaller than unity.\footnote{The situation is less clear for strongly bound orbits around a near-extremal black hole, where one may expect $\omega\sim 1/M$. This case will require further investigation. We suspect that coupling between modes remains subdominant even in that case, thanks to the time-delay effect noted in \cite{Gralla:2016qfw} (recalling that coupling terms all involve $t$ derivatives).}
This can be used to one's advantage in numerical implementations, as we further discuss in Sec.\ \ref{Sec:jumps} below.

The term $\propto \phi^{\ell}_{,tt}$ may be eliminated from Eq.\ (\ref{Teukolsky1+1}) by introducing a modified ($\ell,m,s$-dependent) radial coordinate $\tilde r_*$, satisfying  
\begin{equation}
\frac{d\tilde r_*}{dr_*}=\sqrt{1-4K(r)a^2C_0^{\ell}} =: \beta_{\ell m}(r)\, .
\end{equation}
The expression under the square root here is positive definite, and smaller than $1$, since $0<4Ka^2<1$ and also (it can be shown) $0<C_0^{\ell}<1$ for all relevant $\ell,m,s$.   We note that $\beta\to 1$ in both limits $r\to\infty$ and $r\to r_+$, meaning the modified coordinate $\tilde r_*$ coincides with the usual $r_*$ in both limits. Introducing also the modified advanced and retarded times $\tv:=t+\tilde r_*$ and $\tu:=t-\tilde r_*$, Eq.\ (\ref{Teukolsky1+1}) becomes 
\begin{align} \label{Teukolsky1+1v2}
\phi^\ell_{,\tu\tv} + \tilde U(r)\phi^\ell_{,\tu} + \tilde V(r)\phi^\ell_{,\tv}  + \tilde W(r)\phi^\ell 
\hspace{15mm}
\nonumber\\ 
+\tilde K(r){\cal I}(\phi^{\ell\pm1},\phi^{\ell\pm2})=0,
\end{align}
with
\begin{equation}
\tilde K(r):=K(r)/\beta^2,\quad\quad
\tilde W(r):=W(r)/\beta^2,
\end{equation}
\begin{equation}
\tilde U:=\frac{1}{2\beta^{2}}\left[(1+\beta)U+(1-\beta)V+\frac{1}{2}\frac{d\beta}{dr_*} \right],
\end{equation}
\begin{equation}
\tilde V:=\frac{1}{2\beta^{2}}\left[(1+\beta)V+(1-\beta)U-\frac{1}{2}\frac{d\beta}{dr_*} \right].
\end{equation}
The form (\ref{Teukolsky1+1v2}) (without a $\phi^{\ell}_{,tt}$ term) is more conveniently amenable to a finite-difference representation, especially in double-null-type coordinates.

We envisage a numerical implementation of Eq.\ (\ref{Teukolsky1+1v2}) using a finite-difference scheme based on $\tu,\tv$ coordinates. For stationary modes of the perturbation, and in the Schwarzschild case, the entire second line of (\ref{Teukolsky1+1v2}) vanishes, and the equation takes a simple form, ready for numerical implementation $\ell$ by $\ell$. In the general Kerr case, the equation can first be recast in a matrix form, introducing the vector variable $\vec{\phi}:=\{\phi^{\ell=1},\phi^{\ell=2},\ldots,\phi^{\ell_{\rm max}}\}$, where $\ell_{\rm max}$ is a suitable cutoff; in many problems, including the GSF problem, the large-$\ell$ truncation error may be controlled and made sufficiently small. The matrix equation can then be discretized and solved as in the Schwarzschild case, this time obtaining all $\ell$ modes at once. Since the coupling between modes is finite and ``weak'' (modes couple only to nearest and next-to-nearest neighbors), the matrices involved are band diagonal and hence comfortably amenable to numerical manipulation.

\subsection{Boundary conditions for $\phi^\pm_{s\ell m}$} \label{subsec:BC}

Typically, and in the GSF problem in particular, we require the physical, ``retarded'' solution for the metric perturbation, i.e., the one for which there is no radiation coming in from past null infinity, and no radiation coming out of the past event horizon. In frequency-domain implementations, and also in Cauchy-type time evolutions, this requirement is imposed via a choice of boundary conditions on suitable timelike boundaries.
In a characteristic-type 1+1D evolution of the kind we have in mind here, the numerical domain has no timelike boundaries, and the solution is completely determined once initial data are specified on two initial characteristic rays. In principle, one should be able to select the retarded solution via a suitable choice of characteristic initial data (though in practice we shall adopt a much simpler approach, described in Sec.\ \ref{Sec:numerics} below). But even though boundary conditions are not actively imposed at each time step, knowledge of the form of physical boundary conditions is still important, for a number of reasons.
First, as we shall see, such knowledge informs our choice of regulator functions [specifically, the factor $(r\Delta^s)^{-1}$ in Eq.\ (\ref{expansion})] that control the behavior of the numerical field at large retarded and advanced times. Our choice of regulator will be such that the physical solution is globally bounded in magnitude. Second, once a numerical solution is obtained, it is important to check that it is indeed the desired physical solution. Third, there are cases where pieces of the physical field may be determined analytically (see Sec.\ \ref{m=0} for an example), and in such cases the choice of a particular solution requires knowledge of the physical boundary conditions. 

Our goal now, therefore, is to prescribe physical boundary conditions for the time-radial fields $\phi^{\pm}_{s\ell m}(t,r)$. In Sec.\ \ref{subsec:BCinf} we will consider the behavior at null infinity, and in Sec.\ \ref{subsec:BCEH} the behavior at the event horizon. Section \ref{subsec:BCstat} will discuss the special case of stationary modes. In all cases we will base our analysis on a study of the asymptotic form of solutions to the 1+1D time-domain Teukolsky equation (\ref{Teukolsky1+1v2}). We recall the coordinates $(\tu,\tv)$ are interchangeable with the standard $(u,v)$ in both asymptotic limits.

\subsubsection{Behavior at null infinity}\label{subsec:BCinf}

Consider solutions of (\ref{Teukolsky1+1v2}) that for $r\gg M$ have the asymptotic forms $\sim r^{\alpha}e^{-i\omega u}$ (``outgoing waves'') or $\sim r^{\beta}e^{-i\omega v}$ (``incoming waves''), for some frequency $\omega\ne 0$ (the case $\omega=0$ will be considered separately below). We determine the powers $\alpha$ and $\beta$ by substituting each of these asymptotic-form {\it Ans\"{a}tze} in Eq.\ (\ref{Teukolsky1+1v2}), then expanding in powers of $1/r$ (at fixed $t$), noting $\tilde V(r)=s/r+O(1/r^2)$ and $\tilde K(r),\tilde U(r),\tilde W(r)=O(1/r^2)$. From the leading-order term of each expansion we readily find $\alpha=0$ and $\beta=2s$. Thus we have the two asymptotic solutions 
\begin{eqnarray}\label{AsympInfomega}
\phi^+_{s\ell m\omega}&\sim& e^{-i\omega u} \quad \text{(physical)},
\nonumber\\
\phi^+_{s\ell m\omega}&\sim& r^{2s} e^{-i\omega v}\quad \text{(nonphysical)}.
\end{eqnarray}
It is easy to check, using (\ref{expansion}) and the reconstruction formula (\ref{reconstruction_pm}), that the first solution yields a reconstructed metric with a large-$r$ asymptotic form $h_{\alpha\beta}^{{\rm rec}+}\sim e^{-i\omega u}/r$ (in suitable Cartesian coordinates), representing a purely outgoing wave---hence the designation ``physical''. The second solution yields a perturbation $h_{\alpha\beta}^{{\rm rec}+}\sim e^{-i\omega v}$ (multiplied by some factor of $r$), which does not represent a purely outgoing wave---hence ``nonphysical''. 

We note that, for the physical solution, the magnitude of $\phi_{s\ell m}^+$ approaches a constant, generally nonzero value at future null infinity ($v\to\infty$ for any fixed $u$). This behavior, which is computationally desirable, was achieved by introducing the factor $(r\Delta^s)^{-1}\sim r^{-2s-1}$ in Eq.\ (\ref{expansion}). Without this factor, the physical solution would blow up as $\sim r^3$ for $s=-2$ (IRG), or would fall off rapidly, as $\sim r^{-5}$, for $s=+2$ (ORG), both types of behavior being computationally problematic. 

Note also that, for $s=+2$, any nonphysical solution blows up rapidly (as $\sim r^4$) at infinity. This means that, in an ORG implementation, a numerical solution $\phi_{s\ell m}^+$ that can be demonstrated to remain bounded at infinity (even as the finite-difference step size tends to zero) is automatically guaranteed to be the physical solution, i.e., the one satisfying the correct, outgoing boundary conditions. An IRG calculation does not share this convenient feature: for $s=-2$, nonphysical modes decay rapidly  (as $\sim r^{-4}$) at infinity and would be hard to identify in the data. In this sense, it is computationally advantageous to calculate the external field $\phi_{s\ell m}^+$ in the ORG.

\subsubsection{Behavior at the event horizon}\label{subsec:BCEH}

Moving on to consider the behavior along the horizon, examine the form of (\ref{Teukolsky1+1v2}) for small $\Delta$. Noting $\tilde K(r),\tilde V(r),\tilde W(r)=O(\Delta)$ while $\tilde U(r)=O(\Delta^0)$, we find that, at leading order in $\Delta$, the equation reduces to 
\begin{equation}
\phi^-_{,uv} + \left(\frac{s(M-r_+)}{2Mr_+}+im\Omega_{\rm H}\right)\phi^-_{,u}=0.
\end{equation}
This equation admits the two pure-mode asymptotic solutions 
\begin{eqnarray}\label{AsympEHomega}
\phi^-_{s\ell m\omega}&\sim& e^{-i\omega v} \quad \text{(physical)},
\nonumber\\
\phi^-_{s\ell m\omega}&\sim& \Delta^s e^{-i\omega u}e^{-2i m\Omega_{\rm H}r_*}\quad \text{(nonphysical)}.
\end{eqnarray} 
It can be checked that the first solution yields a reconstructed perturbation that (in suitable, horizon-regular coordinates such as $\{v,r,\theta,\tilde\varphi\}$) has the asymptotic form $h_{\alpha\beta}^{\rm rec-}\sim e^{-i\omega v}$, representing a purely ingoing wave at the future event horizon. This is therefore the ``physical'' solution.\footnote{From Eq.\ (\ref{expansion}) we see that, for the physical solution, the Hertz potential $\Phi_s^-$ itself is $\propto\Delta^{-s}$ at the horizon. This apparent irregular behavior owes itself simply to the irregularity of the Kinnersley tetrad at the horizon; see, e.g., Sec.\ V of \cite{Barack:1999ya}.} For the physical solution, $\phi^-_{s\ell m\omega}$ approaches a finite, generally nonzero value at the horizon ($u\to\infty$ for any fixed $v$), which is computationally desirable. Indeed, to achieve this was the purpose of our introduction of a regulator factor $\Delta^{-s}$ in Eq.\ (\ref{expansion}).

It can also be checked that the second solution in (\ref{AsympEHomega}) produces a reconstructed perturbation that, in coordinates regular on the past event horizon, has the asymptotic behavior $h_{\alpha\beta}^{\rm rec-}\sim e^{-i\omega u}$ there.\footnote{Note that our coordinate $\tilde\varphi$ is {\it not} regular on the past horizon, and it is this coordinate irregularity that gives rise to the singular factor $e^{-2i m\Omega_{\rm H}r_*}$ in the ``nonphysical'' solution in (\ref{AsympEHomega}). When we say that this solution is nonphysical we do not refer to this coordinate irregularity but to the fact that the solution represents outgoing waves at the past horizon.} This solution thus represents nonphysical outgoing waves at the past horizon. Note that for $s=-2$ the nonphysical modes blow up as $\sim \Delta^{-2}$ at the horizon. Hence, in an IRG reconstruction it is sufficient to demonstrate the boundedness of the internal numerical solution $\phi^-_{s\ell m\omega}$ at the horizon in order to establish that it represents the true, physical solution (up to numerical error). In the ORG, on the other hand, nonphysical modes decay rapidly (as $\sim \Delta^2$) near the horizon and would be hard to identify there if they existed in the numerical data.  Thus, there is a computational advantage in calculating the internal field $\phi_{s\ell m}^-$ in the IRG.



\subsubsection{Stationary modes}\label{subsec:BCstat}

Finally, let us consider the asymptotic behavior of stationary, $t$-independent (or, equivalently, $\omega=0$) modes. Substituting the {\it Ansatz} $\phi\sim r^{\alpha}$ in Eq.\ (\ref{Teukolsky1+1v2}) and considering the leading-order beavior at $r\gg M$, one obtains $\alpha=s-\ell$ or $\ell+s+1$, and hence the two asymptotic solutions 
\begin{eqnarray}\label{AsympInfstat}
\phi^+_{s\ell m}&\sim& r^{-\ell+s} \quad \text{(physical)},
\nonumber\\
\phi^+_{s\ell m}&\sim& r^{\ell+s+1} \quad \text{(nonphysical)}.
\end{eqnarray}
The designations `physical' and `nonphysical' here come from examining the behavior of the physical perturbation associated with each solution. To eliminate gauge dependence, it is instructive to consider, for example, the corresponding Weyl scalars $\Psi_0$ or $\Psi_4$. Using Eq.\ (\ref{expansion}) with (\ref{invertion_rad_IRG}) and (\ref{invertion_rad_ORG}) [or with (\ref{invertion_ang_IRG}) and \ref{invertion_ang_ORG})] shows $\Psi_0,\Psi_4\sim r^{-\ell-3}$ for the first solution in (\ref{AsympInfstat}), and  $\Psi_0,\Psi_4\sim r^{\ell-2}$ for the second, in both the IRG and ORG cases. Thus, the first solution corresponds to a perturbation whose curvature decays at infinity (hence `physical'), while the second solution (`nonphysical') corresponds to a perturbation whose curvature does not decay at infinity (and, for $\ell>2$, it actually blows up there). 

We move on to consider the behavior on the event horizon. Two stationary asymptotic solutions there are  
\begin{eqnarray}\label{AsympEHstat}
\phi^-_{s\ell m}&\sim& {\rm const} \quad \text{(physical)},
\nonumber\\
\phi^-_{s\ell m}&\sim& \Delta^s \quad \text{(nonphysical)},
\end{eqnarray}
where, importantly, the first solution has a regular Taylor expansion on the horizon, while the second solution also contains a high-order contribution of the form $(\log\Delta)\times$ a Taylor series in $\Delta$. It can be checked that the corresponding Weyl scalars have the behavior $\Psi_s\sim \Delta^{-s}$ for the first solution, and $\Psi_s\sim \Delta^0$ (plus higher-order log terms) for the second solution. As explained (e.g.) in Sec.\ V of Ref.\ \cite{Barack:1999ya}, carefully taking into account the irregularity of the Kinnersley tetrad at the horizon, for a smooth physical perturbation it is not the Weyl scalars themselves that are regular (smooth) at the horizon, but rather the product $\Delta^s\Psi_s$. Applying this criterion to our solutions, we have that $\Delta^s\Psi_s$ is perfectly smooth for the first solution (``physical''),  but non-smooth for the second solution (``nonphysical''). In the latter case, $\Delta^s\Psi_s$ blows up like $\Delta^{-2}$ for $s=-2$, while for $s=+2$ the differentiability is spoiled by the $\log\Delta$ term.

We note that, in terms of our variable $\phi^\pm(t,r)$, the stationary piece of the physical perturbation is bounded everywhere, just like the rest of the perturbation. For nonphysical stationary perturbations, $\phi^-$ blows up on the horizon in the IRG case, and $\phi^+$ blows up at infinity in both the IRG and ORG cases. This is true for all relevant values of $\ell$, i.e.\ $\ell\geq 2$.

\subsubsection{Summary}\label{subsec:BCsum}

In summary, our choice of time-radial fields $\phi^\pm_{s\ell m}(t,r)$ is such that, in terms of these variables, the physical, retarded solution is bounded both at infinity and on the horizon---and, in fact, anywhere else in the computation domain. This is a convenient feature, computationally. Furthermore, we have noted that all nonphysical ORG solutions $\phi^+_{s\ell m}$ blow up at infinity, while all nonphysical IRG solutions $\phi^-_{s\ell m}$ blow up at the horizon. Thus, boundedness of an ORG solution at infinity implies that the correct outgoing boundary conditions are satisfied at infinity, and boundedness of an IRG solution at the horizon implies that the correct ingoing boundary conditions are satisfied on the horizon.

Unfortunately, we cannot make a stronger statement: In the ORG case, boundedness of the solution on the horizon does {\em not} necessarily mean that boundary conditions are satisfied there, because nonphysical ORG modes are subdominant on the horizon. Similarly, in the IRG case, boundedness of the solution at infinity does {\em not} necessarily imply, in general, that boundary conditions are satisfied there, because nonphysical IRG modes are subdominant at infinity (with the exception of stationary modes, which blow up there). This situation suggests that, at least from a computational point of view, it would be convenient to work with a mixed-gauge field composed of the ORG $\phi^+_{s\ell m}$ and the IRG $\phi^-_{s\ell m}$. However, the formulation of jump conditions on the orbit (see below) would then be harder, and one would also need to generalize the self-force formulation to accommodate the possibility of such a mixed-gauge perturbation. It would be worth exploring the mixed-gauge idea in future work, but here we shall stick with the more straightforward single-gauge approach.

\subsection{Jump conditions for $\phi^{\pm}$ across $\cal S$}\label{Sec:jumps}

Finally, we need a set of rules that relate the fields $\phi^+_{s\ell m}$ and $\phi^-_{s\ell m}$ along the particle's timelike trajectory in the 1+1D domain. Specifically, we need the ``jumps''
\begin{equation}
\J{\phi}:=\lim_{\epsilon\to 0} \left[\phi^+(t,r_p(t)+\epsilon)-\phi^-(t,r_p(t)-\epsilon)\right],
\end{equation}
as well as the jumps in the first derivatives---say, $\J{\phi_{,\tu}}$ and $\J{\phi_{,\tv}}$. (In this subsection we occasionally, where possible, omit the indices $s\ell m$ for brevity.) The jumps in higher-order derivatives may also be required, depending on the particular numerical method implemented and on the order of numerical convergence sought. Our goal now is to describe a method for determining these jumps, as functions along the particle's worldline. We will assume that the jumps $\J{\psi_{\pm2}}$ in the physical Weyl scalars, and in their derivatives, are already known. These jumps may be deduced directly, in analytic form, from the source term of the (1+1D version of the) Teukolsky equation, without needing to solve the equation for the Weyl scalars themselves---a specific example will be worked out explicitly in Appendix \ref{AppA}. Thus, we will be seeking to determine the jumps in the Hertz potential and its derivatives in terms of the jumps in the physical Weyl scalars and their derivatives.

Our starting point is Eqs.\ (\ref{invertion_rad_IRG}), (\ref{invertion_rad_ORG}), (\ref{invertion_ang_IRG}) and (\ref{invertion_ang_ORG}), as applied to the vacuum solutions $\Phi_s^{\pm}$. Recall these are relations that the vacuum Hertz potential must satisfy given the physical Weyl scalars $\Psi_0$ or $\Psi_4$. First, we need to obtain the 1+1D version of these equations. To this end, we substitute the expansion (\ref{expansion}) for $\Phi^{\pm}_s$ on the left-hand side of each of these relations, and on the right-hand side we substitute for $\Psi_4=:\varrho^4\Psi_{s=-2}$ and  $\Psi_0=:\Psi_{s=+2}$ using the analogous expansions 
\begin{equation} \label{expansionWeyl}
\Psi_s^{\pm}=
(r\Delta^{s})^{-1}
       \sum_{\ell=2}^{\infty}\sum_{m=-\ell}^{\ell}
       \psi^{\pm}_{s\ell m}(t,r) {}_s\!Y_{\ell m}(\theta,\tilde\varphi)
\end{equation}
for $s=\pm 2$.

Considering first the relations (\ref{invertion_rad_IRG}) and (\ref{invertion_rad_ORG}) (whose frequency-domain versions are often referred to as ``radial inversion formulas''), we observe that the differential operators on the left-hand side do not couple between different $\ell$ modes: the equations relate each (spin-weighted spherical-harmonic) $\ell$ mode of the Hertz potentials $\bar\Phi^{\pm}_s$ and their derivatives to the same $\ell$ mode of the Weyl scalars $\Psi^{\pm}_{-s}$ (with $m$ opposite in sign). Using (\ref{symmetry}) and the orthogonality of ${}_s\!Y_{\ell m}$ we obtain, for each $\ell,m$,
\begin{equation} \label{1+1IRGinversion}
8r\Delta^2{\cal D}_l^4 \left(\Delta^2\bar\phi^{{\rm IRG}\pm}_{\ell m}/r\right)
= (-1)^m\psi^{\pm}_{2,\ell,-m},
\end{equation}
\begin{equation} \label{1+1ORGinversion}
\frac{1}{2}r\tilde{\cal D}_n^4 \left(\bar\phi^{{\rm ORG}\pm}_{\ell m}/r\right)
= (-1)^m\psi^{\pm}_{-2,\ell,-m},
\end{equation}
where the differential operators are
\begin{eqnarray} \label{1+1IRGinversionOperator}
{\cal D}_l&:=&\Delta^{-1}\left[(r^2+a^2)\partial_v-ima\right],
\nonumber\\
\tilde{\cal D}_n&:=&-\Delta^{-1}(r^2+a^2)\partial_u;
\end{eqnarray}
here $\partial_v$ is taken with fixed $u$, and $\partial_u$ is taken with fixed $v$.  We observe that (\ref{1+1IRGinversion}) and (\ref{1+1ORGinversion}) are effectively {\em ordinary} differential equations for $\bar\phi^{{\rm IRG}\pm}_{\ell m}$ and $\bar\phi^{{\rm ORG}\pm}_{\ell m}$, with independent variables $v$ and $u$, respectively.

In general, the alternative relations (\ref{invertion_ang_IRG}) and (\ref{invertion_ang_ORG}) (``angular inversion formulas'') are less useful here, because they feature operators that
{\it do} mix between different $\ell$ modes, giving rise to mode-coupled relations; even worse, the equations relate each $\ell$ mode of $\Phi^{\pm}$ to an {\em infinite} number of $\ell$ modes of $\Psi_s^{\pm}$. The coupling disappears only in the Schwarzschild case, $a=0$, where, in fact, the angular inversion formulas become extremely simple (since the operators $\tilde{\cal L}_s$ and ${\cal L}_s$ reduce to, respectively, spin-lowering and spin-raising operators when they act on spin-weighted spherical harmonics). This simplicity was noted previously by Lousto and Whiting in \cite{Lousto:2002em}, where they considered 1+1D metric reconstruction in Schwarzschild. Here, however, our ambition is to treat the more general Kerr case, so we will utilize the ``radial'' inversion equations---even as (in the next two sections) we consider a Schwarzschild example. 

Focusing thus on the 1+1D inversion formulas (\ref{1+1IRGinversion}) and (\ref{1+1ORGinversion}), we first note that the {\em general} solution for each of these two ODEs can be written down in a simple closed form (involving four nested integrals with respect to $v$ or $u$, respectively). 
Based on the form of these general solutions it is straightforward to show (in analogy with Ori's analysis in \cite{Ori:2002uv}) that $\phi^{{\rm IRG}\pm}_{\ell m}$ and $\phi^{{\rm ORG}\pm}_{\ell m}$ each admits a {\em unique} particular solution that also satisfies the vacuum Teukolsky equation (\ref{Teukolsky1+1v2}), as required. This confirms the uniqueness of the Hertz potential in the reconstruction procedure.

However, the above closed-form particular solutions for $\phi^{{\rm IRG}\pm}_{\ell m}$ and $\phi^{{\rm ORG}\pm}_{\ell m}$ involve integrals of the Weyl scalars ($\psi_2^{\pm}$ and $\psi_{-2}^{\pm}$, respectively) along rays extending to infinity and down to the event horizon, and to evaluate them in practice would require solving the appropriate sourced Teukolsky equations as a preliminary step. In our method we wish to bypass this preliminary step, and work directly with the Hertz potential; we wish to have no recourse to knowledge of the Weyl scalars themselves (except the values of their jumps across the particle, which, as mentioned, are accessible directly from the source of the Teukolsky equation). Our goal, therefore, is to express the jumps in $\phi$ (and its derivatives) in terms of the {\em jumps} in $\psi$ (and its derivatives) alone. In what follows we describe a procedure that achieves that.

Let us start with the IRG case (the ORG case will follow analogously). Evaluating the difference between the `$+$' and `$-$' versions of Eq.\ (\ref{1+1IRGinversion}) at $r=r_{\rm p}(t)$ yields a relation of the form 
\begin{equation}\label{inversion_form}
\sum_{n=0}^4 f_n(r_p)[\partial_{\tv}^n \phi^{{\rm IRG}}_{\ell m}]=(-1)^m\J{\bar\psi_{2,\ell,-m}},
\end{equation}
in which $f_n(r_{\rm p})$ are some smooth functions along the orbit, and the jumps on the right-hand side are assumed known. For a reason that will become clear shortly, we also need the $\tv$ derivative of (\ref{1+1IRGinversion}), which yields
\begin{equation}\label{inversionv_form}
\sum_{n=0}^5 \tilde f_n(r_p)[\partial_{\tv}^n \phi^{{\rm IRG}}_{\ell m}]=(-1)^m\J{(\bar\psi_{2,\ell,-m})_{,\tv}},
\end{equation}
where  $\tilde f_n(r_{\rm p})$ are some other smooth functions, and the jumps on the right-hand side are also known. Our goal is to determine the jumps $\J{\phi}$ and $\J{\phi_{,\tv}}$, as well as $\J{\phi_{,\tu}}$, as functions along the orbit (here and in the following discussion we omit the labels $\ell,m$ and IRG for brevity). 

Using an overdot to denote $d/d\tau$, where $\tau$ is proper time along the orbit, we write
\begin{eqnarray} \label{phidot}
\dot{\J{\phi}}&=&\dot \tu_{\rm p}\J{\phi_{,\tu}}+\dot \tv_{\rm p}\J{\phi_{,\tv}} , \\
\label{phivdot}
[\dot\phi_{,\tv}]&=&\dot \tu_{\rm p}\J{\phi_{,\tv\tu}}+\dot \tv_{\rm p}\J{\phi_{,\tv\tv}} ,
\end{eqnarray}
where $\tu_{\rm p}(\tau)$ and $\tv_{\rm p}(\tau)$ are the values of $\tu$ and $\tv$  at a worldline point with proper time $\tau$. Equation (\ref{phidot}) gives $\J{\phi_{,\tu}}$ in terms of $\dot{\J{\phi}}$ and $\J{\phi_{,\tv}}$. In Eq.\ (\ref{phivdot}) we replace $\J{\phi_{,\tv\tu}}$ in favor of $\J{\phi}$, $\J{\phi_{,\tv}}$ and $\J{\phi_{,\tu}}$ using the Teukolsky equation (\ref{Teukolsky1+1v2}), and hence express $\J{\phi_{,\tv\tv}}$ in terms of $\J{\phi}$ and $\J{\phi_{,\tv}}$ (and their $\tau$ derivatives) alone. (For $a\ne 0$ this relation will involve also the jump in the coupling terms $\cal I$. Let us ignore these terms for a moment to simplify the discussion; we shall return to them momentarily.)  Next, we write
\begin{eqnarray}
\ddot{\J{\phi}}&=&\ddot{\tv}_{\rm p}\J{\phi_{,\tv}}+\ddot{\tu}_{\rm p}\J{\phi_{,\tu}}
+\dot{\tu}_{\rm p}^2\J{\phi_{,\tu\tu}}+\dot{\tv}_{\rm p}^2\J{\phi_{,\tv\tv}}\nonumber\\
&&+2\dot{\tv}_{\rm p}\dot{\tu}_{\rm p}\J{\phi_{,\tv\tu}},
\\
{}[\ddot\phi_{,\tv}]&=&\ddot{\tv}_{\rm p}\J{\phi_{,\tv\tv}}+\ddot{\tu}_{\rm p}\J{\phi_{,\tv\tu}}+\dot{\tu}_{\rm p}^2\J{\phi_{,\tv\tu\tu}}+\dot{\tv}_{\rm p}^2\J{\phi_{,\tv\tv\tv}}\nonumber\\
&&+2\dot{\tv}_{\rm p}\dot{\tu}_{\rm p}
\J{\phi_{,\tv\tv\tu}},
\end{eqnarray}
and use (\ref{Teukolsky1+1v2}) again to replace all mixed-derivative jumps with lower-order-derivative jumps. We thus express $\J{\phi_{,\tu\tu}}$ and (in turn) $\J{\phi_{,\tv\tv\tv}}$ in terms of $\J{\phi}$ and $\J{\phi_{,\tv}}$ (and their first and second $\tau$ derivatives) alone. 

Proceeding in a similar way, we can determine  $\J{\phi_{,\tu\tu\tu}}$ and $\J{\phi_{,\tv\tv\tv\tv}}$ in terms of $\J{\phi}$, $\J{\phi_{,\tv}}$ and their first, second and third $\tau$ derivatives; and finally we can determine $\J{\phi_{,\tu\tu\tu\tu}}$ and $\J{\phi_{,\tv\tv\tv\tv\tv}}$ in terms of $\J{\phi}$, $\J{\phi_{,\tv}}$ and their first, second, third and fourth $\tau$ derivatives. Equations (\ref{inversion_form}) and (\ref{inversionv_form}) can thus be written as a coupled set of ODEs for $\J{\phi}$ and $\J{\phi_{,\tv}}$:
\begin{align}\label{JumpEq1}
\sum_{n=0}^3 \left(a_n(\tau)\frac{d^n\J{\phi}}{d\tau^n} + b_n(\tau)\frac{d^n\J{\phi_{,\tv}}}{d\tau^n}\right)
+\text{$\cal I$ terms}
\nonumber\\
=
(-1)^m\J{\bar\psi_{2,\ell,-m}},
\end{align}
\begin{align}\label{JumpEq2}
\sum_{n=0}^4 \left(c_n(\tau)\frac{d^n\J{\phi}}{d\tau^n} + d_n(\tau)\frac{d^n\J{\phi_{,\tv}}}{d\tau^n}\right)
+\text{$\cal I$ terms}
\nonumber\\
=
(-1)^m\J{(\bar\psi_{2,\ell,-m})_{,\tv}},
\end{align}
where $a_n(\tau),\ldots,d_n(\tau)$ are certain smooth functions along the worldline (depending only on $r_{\rm p},\dot{r}_{\rm p},\ldots , d^4r_{\rm p}/d\tau^4$, as well as on $\ell$). The terms collected under `${\cal I}$ terms' are certain linear combinations of the coupling term ${\cal I}(\phi^{\ell\pm 1},\phi^{\ell\pm 2})$ and its $v$ and $u$ derivatives (up to third derivatives), which have entered the relations via our use of the Teukolsky equation (\ref{Teukolsky1+1v2}).
Unfortunately, the general explicit form of Eqs.\ (\ref{JumpEq1}) and (\ref{JumpEq2}) is too unwieldy to be presented here in any meaningful way, but it can be straightforwardly obtained using computer algebra tools, following the procedure described above. In Sec.\ \ref{Sec:example} we will present explicit expressions for the special case of circular geodesic orbits in Schwarzschild spacetime [where all ${\cal I}$ terms drop, and the coefficients $a_n,\ldots,d_n$ depend only on $r_{\rm p}$(=const)].  

Let us for the moment ignore the coupling terms in Eqs.\ (\ref{JumpEq1}) and (\ref{JumpEq2}). Then these equations constitute a coupled set of fourth-order ODEs for the jumps $\J{\phi_{\ell m}}$ and $\J{\phi_{\ell m,\tv}}$ as functions along the orbit. How to solve these equations depends on the particular problem under consideration. If the perturbation is sourced by a particle on a fixed bound geodesic orbit, then the jumps may be assumed to exhibit the same periodicity as the orbit (two fundamental frequencies, in general), and the ODEs  (\ref{JumpEq1}) and (\ref{JumpEq2}) can then be conveniently recast as a set of algebraic equations, one for each frequency (the assumption of periodicity then effectively selects a particular solution of the ODEs).\footnote{By solving the jump equations (\ref{JumpEq1}) and (\ref{JumpEq2}) frequency by frequency we would {\em not} be reverting to the standard frequency-domain approach to metric reconstruction: the field equation (\ref{Teukolsky1+1}) would still be solved in the time domain. } If the perturbation is sourced by a slowly evolving orbit (e.g., under the effect of the self-force), then one should still be able to obtain a frequency-by-frequency algebraic solution at some initial point along the orbit, then solve the set (\ref{JumpEq1}) and (\ref{JumpEq2}) as ODEs, starting from these initial conditions. For nonperiodic (parabolic- or hyperbolic-type) orbits, initial conditions may be formulated at infinity, using an asymptotic analysis. In Sec.\ \ref{Sec:example} we give the explicit physical solution of Eqs.\ (\ref{JumpEq1}) and (\ref{JumpEq2}) for the special case of circular geodesic orbits in Schwarzschild spacetime.

How should the $\ell$-mode coupling terms in Eqs.\ (\ref{JumpEq1}) and (\ref{JumpEq2}) be dealt with, in the Kerr case? As already mentioned, we expect the $\cal I$ term in the field equation (\ref{Teukolsky1+1}) to be small, in general, in a certain relative sense. The coupling terms in the jump equations (\ref{JumpEq1}) and (\ref{JumpEq2})  will be small in the same sense. One could then incorporate these terms perturbatively, using an iterative scheme: In the first iteration, Eqs.\ (\ref{JumpEq1}) and (\ref{JumpEq2}) are solved for each $\ell,m$ with the coupling terms dropped. The solutions are then used to calculate the $\cal I$ terms in Eqs.\ (\ref{JumpEq1}) and (\ref{JumpEq2}), and the equations are solved again, with these $\cal I$ terms as sources. One keeps iterating in this manner until a sufficiently convergent solution is achieved; the smaller the coupling terms are in relative  magnitude, the less iterations should be required. How computationally tasking this procedure may prove to be would depend on the number of iterations required and on whether the jump equations are solved as ODEs or via a mode decomposition. Yet we expect the computational cost of calculating the jumps to be negligible compared with the cost of solving the field equation in the time domain.

At any rate, once the jumps $\J{\phi}$ and $\J{\phi_{,\tv}}$ have been computed as functions along the orbit, the jumps in any higher $\tv$ and $\tu$ derivatives of $\phi$ are calculable algebraically using the order-reduction procedure described a couple of paragraphs above: $\J{\phi_{,\tu}}$ is obtained from (\ref{phidot}), then $\J{\phi_{,\tv\tv}}$ is obtained from (\ref{phivdot}), and so on.

Finally, the ORG version of the problem is dealt with in a completely analogous manner, this time starting with Eq.\ (\ref{1+1ORGinversion}). The procedure again yields ODEs of the form (\ref{JumpEq1})-(\ref{JumpEq2}), but now with $\J{\phi_{,\tv}}$ replaced with $\J{\phi_{,\tu}}$ and, on the right-hand side, $\bar\psi_{2,\ell,-m}$ replaced with $\bar\psi_{-2,\ell,-m}$. The explicit form of the coefficients $a_n(\tau),\ldots,d_n(\tau)$ and of the $\cal I$ terms will also differ.  

\subsection{Summary of proposed evolution scheme}

In summary, our evolution scheme for the Hertz potential consists of the 1+1D evolution equation (\ref{Teukolsky1+1v2}), the asymptotic boundary conditions (\ref{AsympInfomega}) and (\ref{AsympEHomega}) (``physical''), and jump conditions across the particle's orbits, given as solutions to Eqs.\ (\ref{JumpEq1}) and (\ref{JumpEq2}). We have sketched how the actual jump conditions for the modes of the Hertz potential may be obtained in practice, depending on the orbital configuration.  

In the next two sections we will illustrate the application of our method with the concrete example of a particle moving in a fixed circular geodesic orbit around a Schwarzschild black hole. Section \ref{Sec:example} will formulate the evolution problem as applied to this case, including explicit expressions for the jumps across the particle. Section \ref{Sec:numerics} will present a numerical implementation. For concreteness and brevity, we will concentrate on the IRG problem.

\section{circular orbits in Schwarzschild spacetime}\label{Sec:example}

Specialized to $a=0$ (Schwarzschild case) and $s=-2$ (IRG),
the 1+1D Teukolsky equation (\ref{Teukolsky1+1v2}) becomes
\begin{align} \label{Teukolsky1+1Sch}
\phi^\ell_{,uv} + U(r)\phi^\ell_{,u} + V(r)\phi^\ell_{,v}  + W(r)\phi^\ell =0,
\end{align}
with
\begin{equation} \label{UV}
U(r)=\frac{2M}{r^2},\quad\quad
V(r)=-\frac{2f}{r},
\end{equation}
\begin{align} \label{W_Sch}
W(r)=\frac{f}{4}\left(
\frac{\lambda}{r^2}-\frac{2M}{r^3}\right),
\end{align}
where we have introduced 
\begin{equation}
f(r):=1-2M/r,
\quad\quad
\lambda:=(\ell+2)(\ell-1) .
\end{equation}
Note that, in the Schwarzschild case, (i) the 1+1D Teukolsky equation does not couple between $\ell$ modes, so individual modes evolve independently of each other; (ii) the equation has no reference to the azimuthal number $m$ (as expected, by virtue of the background's spherical symmetry); and (iii) the modified coordinates $\tu,\tv$ reduce to the standard coordinates $u,v$. 

Next consider the jump equations (\ref{JumpEq1})-(\ref{JumpEq2}). For $a=0$, the coupling terms drop. Specializing further to a circular geodesic orbit with radius $r_{\rm p}=r_0(={\rm const}>2M)$, the coefficients on the left-hand side of (\ref{JumpEq1})-(\ref{JumpEq2}) work out to be 
\begin{eqnarray}
a_0&=&\frac{1}{2}r_0^4f_0^2 \lambda(\lambda+2) \nonumber\\
a_1&=&2r_0^5\left[\lambda-y(\lambda-3)-2y^2(\lambda+5)\right]/\gamma_0, \nonumber\\
a_2&=& 2r_0^6f_0(\lambda+6y)/\gamma_0^2, \nonumber\\
a_3&=& 16Mr_0^6/\gamma_0^3,
\end{eqnarray}
\begin{eqnarray}
b_0&=&0 = b_2, \nonumber\\ 
b_1&=&4r_0^6\left[\lambda-2y(\lambda-1)-6y^2\right]/\gamma_0, \nonumber\\
b_3&=& 8r_0^8/\gamma_0^3,
\end{eqnarray}
\begin{eqnarray}
c_0&=&r_0^3f_0^2(1-y)\lambda(\lambda+2),
\nonumber\\
c_1&=&{f_0 r_0^{4}} \left[
\lambda  (\lambda +5)
-2 (\lambda ^2+2 \lambda -6) y
-2 (4 \lambda +17) y^2  \right.
\nonumber\\ &&
+\left. 12 y^3\right]/\gamma_0 ,
\nonumber\\
c_2&=&2 r_0^5 \left[3 \lambda +(15-7 \lambda)y +2 (\lambda -23) y^2+ 24y^3 \right]/{\gamma_0^2}, \nonumber\\
c_3&=&{2  f_0 r_0^6  (\lambda +22y )}/{\gamma_0^3},\nonumber\\
c_4&=&{16 M r_0^6}/{\gamma_0^4} , \nonumber\\
\end{eqnarray}
\begin{eqnarray}
d_0&=&\frac{1}{2}r_0^4f_0^2\lambda(\lambda+2),\nonumber\\
d_1&=& 2r_0^5f_0\left[3\lambda-5(\lambda-1)y-12y^2\right]/\gamma_0,\nonumber\\
d_2&=&{2 r_0^6 \left[3 \lambda +2 (5-3 \lambda ) y -24 y^2\right]}/{\gamma_0^2} , \nonumber\\
d_3&=&{16 f_0 r_0^7}/{\gamma_0^3} , \nonumber\\
d_4&=&{8 r_0^8}/{\gamma_0^4} , \nonumber\\
\end{eqnarray}
where 
\begin{equation}
y:=\frac{M}{r_0}, \quad\quad
f_0:=f(r_0), \quad\quad
\gamma_0:= \left(1-\frac{3M}{r_0}\right)^{-1/2} .
\end{equation}
We have used here the fact that, for circular geodesics, $\dot{u}_{\rm p}=\dot{v}_{\rm p}(=\gamma_0)$
and all higher-order $\tau$ derivatives of $u_{\rm p}$ and $v_{\rm p}$ vanish.

Furthermore, for a circular geodesic orbit we may assume that $[\phi_{\ell m}]$ depends on time solely via $e^{-im\Omega t}$, where
\begin{equation}\label{Omega}
\Omega:=d\varphi_{\rm p}/dt=\sqrt{M/r_0^3}
\end{equation}
is the orbital angular velocity. Hence, in Eqs.\  (\ref{JumpEq1})-(\ref{JumpEq2}) $d/d\tau$ may be replaced with $-im\Omega(dt_{\rm p}/d\tau)=-im\Omega\gamma_0$. These equations then become algebraic, taking the simple form 
\begin{eqnarray}\label{JumpEqAlgeb}
a_\Sigma\J{\phi} + b_\Sigma\J{\phi_{,v}}&=&
(-1)^m\J{\bar\psi_{2,\ell,-m}}, \nonumber\\
c_\Sigma\J{\phi} + d_\Sigma\J{\phi_{,v}}&=&
(-1)^m\J{\partial_v\bar\psi_{2,\ell,-m}},
\end{eqnarray}
where $a_\Sigma=\sum_{n=0}^3 (-im\Omega\gamma_0)^n a_n $ and similarly for $b_\Sigma$, $c_\Sigma$ and $d_\Sigma$. The solutions are
\begin{eqnarray}\label{Jphi}
\!\!\!\!\!\!\!\!\!
\J{\phi^{\rm IRG}_{\ell m}}&=&\frac{(-1)^m}{\tilde\Delta} \left(d_\Sigma\J{\bar\psi_{2,\ell,-m}}-b_\Sigma\J{\partial_v\bar\psi_{2,\ell,-m}} \right)\!,
\\ \!\!\!\!\!\!\!\!\!
\J{\partial_v\phi^{\rm IRG}_{\ell m}}&=&\frac{(-1)^m}{ \tilde\Delta} \left(a_\Sigma\J{\partial_v\bar\psi_{2,\ell,-m}}-c_\Sigma\J{\bar\psi_{2,\ell,-m}} \right)\!,
\label{Jphiv}
\end{eqnarray}
where $\tilde\Delta:=a_\Sigma d_\Sigma-b_\Sigma c_\Sigma$, and we have restored all labels. We find
\begin{equation}
\tilde\Delta = \frac{1}{4}f_0^4 r_0^8\left[
\lambda^2(\lambda+2)^2+(12mM\Omega)^2 
\right],
\end{equation}
which, we note, is positive definite.

Equations (\ref{Jphi}) and (\ref{Jphiv}) give the desired jumps in the Hertz potential and its $v$ derivative in terms of the jumps in the Weyl scalar ($\Psi_0$) corresponding to the physical perturbation. The latter jumps are easily obtained from the distributional source of the Teukolsky equation satisfied by the Weyl scalar. In Appendix \ref{AppA} we show how the jumps $\J{\partial_v\bar\psi_{2,\ell,-m}}$ and $\J{\bar\psi_{2,\ell,-m}}$ are derived, and give them explicitly for circular geodesic orbits in Schwarzschild.  Once $\J{\phi}$ and $\J{\phi_{,v}}$ are known, the jumps in other derivatives of $\phi$ can be obtained iteratively, by means of the procedure described above. For instance,
\begin{eqnarray}
\J{\phi_{,u}}&=&
-im\Omega \J{\phi}-\J{\phi_{,v}},
\nonumber\\
\J{\phi_{,vu}}&=&- U(r_0)\J{\phi_{,u}} - V(r_0)\J{\phi_{,v}}- W(r_0)\J{\phi} ,
\nonumber\\
\J{\phi_{,uu}}&=&
-im\Omega \J{\phi_{,u}}-\J{\phi_{,vu}},
\nonumber\\
\J{\phi_{,vv}}&=&
-im\Omega \J{\phi_{,v}}-\J{\phi_{,vu}},
\end{eqnarray}
and so on.

\subsection{Analytical solutions for $m=0$}\label{m=0}

Axisymmetric modes (those with $m=0$) are also stationary, and admit simple analytic solutions. We write these solutions here explicitly, as they will be useful in testing our numerical implementation in the next section.  

For $m=0$ we have $\phi^\ell_{,t}=0$, and the homogeneous Teukolsky equation (\ref{Teukolsky1+1Sch}) reduces to 
\begin{equation}\label{eq:TeukolskyStat}
r \left[\left(r^3 f^2 \phi^\ell\right)'/(r^2 f)\right]'-\lambda\,\phi^{\ell}= 0,
\end{equation}
where a prime denotes $d/dr$.
Two linearly independent solutions are
\begin{eqnarray}\label{phiPQ}
\phi^{\ell}&=& \frac{{\sf P}_{\ell}^{m=2}(x)}{\sqrt{\lambda(\lambda+2)}(r-2M)} =:\phi^{\ell}_{P}(r),
\nonumber\\
\phi^{\ell}&=&\frac{{\sf Q}_{\ell}^{m=2}(x)}{\sqrt{\lambda(\lambda+2)}(r-2M)}:=\phi^{\ell}_{Q}(r),
\end{eqnarray}
where ${\sf P}_\ell^{m}$ and ${\sf Q}_\ell^{m}$ are associated Legendre functions of the first and second kinds, respectively, $x:=(r-M)/M$, and the solutions have been normalized so as to render the Wronskian $\ell$ independent: 
\begin{equation}\label{Wron}
(\phi^{\ell}_{P})'\phi^{\ell}_{Q}-\phi^{\ell}_{P}(\phi^{\ell}_{Q})'=\frac{M}{r^4 f^3}.
\end{equation}
The solution $\phi^{\ell}_{P}$ blows up as $\sim r^{\ell-1}$ at infinity but has a regular Taylor expansion at the event horizon. The solution $\phi^{\ell}_{Q}$ is regular at infinity (where it falls off as $\sim r^{-\ell-2}$) but blows up as $\sim f^{-2}$ at the horizon. 

Recalling Eqs.\ (\ref{AsympInfstat}) and (\ref{AsympEHstat}), we see that $\phi^{\ell}_{P}$ satisfies physical boundary conditions on the horizon (but not at infinity), while $\phi^{\ell}_{Q}$ satisfies physical boundary conditions at infinity (but not on the horizon). Therefore, a unique physical solution is given by
\begin{eqnarray}\label{phiStat}
\phi^{-}_{{-2}\ell 0}&=& C^-_{\ell}(r_0)\phi^{\ell}_{P}(r),
\nonumber\\
\phi^+_{{-2}\ell 0}&=& C^+_{\ell}(r_0)\phi^{\ell}_{Q}(r).
\end{eqnarray}
The coefficients $C^\pm_{\ell}(r_0)$ are determined from the two jump conditions $\phi^+_{\ell 0}(r_0)-\phi^-_{\ell 0}(r_0)=\J{\phi_{\ell 0}}$ and $(\phi^+_{\ell 0})'(r_0)-(\phi^-_{\ell 0})'(r_0)=\J{\phi'_{\ell 0}}$, giving
\begin{eqnarray}\label{Cpm}
C^-_{\ell}&=&r_0^4f_0^3\left(\J{\phi_{\ell0}}\phi'_Q(r_0)-\J{\phi_{\ell0}'}\phi^{\ell}_{Q}(r_0)\right)/M,
\nonumber\\
C^+_{\ell}&=&r_0^4f_0^3\left(\J{\phi_{\ell0}}\phi'_P(r_0)-\J{\phi_{\ell0}'}\phi^{\ell}_{P}(r_0)\right)/M,
\end{eqnarray}
where we have substituted for the Wronskian from Eq.\ (\ref{Wron}). The jumps $\J{\phi_{\ell 0}}$ and $\J{\phi'_{\ell 0}}=(2/f_0)\J{\phi_{\ell 0,v}}$ are calculated using Eq.\ (\ref{Jphi}) with Eqs.\ (\ref{Jpsi}) and (\ref{Jpsir}). For $m=0$ we obtain the simple expressions
\begin{eqnarray}\label{JphiStat}
\J{\phi_{\ell 0}}&=& \frac{16\pi\mu\gamma_0}{r_0^2 f_0^2 \lambda(\lambda+2)}
\left(y^2 {\cal Y}^\ell
+i f_0 r_0\Omega{\cal Y}^\ell_{\theta}\right),
\\ \label{JphirStat}
\J{\phi'_{\ell 0}}&=&-\frac{8\pi\mu\gamma_0}{r_0^3 f_0^3 \lambda(\lambda+2)}
\left\{
\left[2f_0(1-y)^2+4y^2+yf_0\lambda\right]{\cal Y}^\ell
\right.
\nonumber\\
&&
\left. +2if_0(1+2y)r_0\Omega{\cal Y}^\ell_{\theta}
-f_0^2 {\cal Y}^\ell_{\theta\theta}
\right\},
\end{eqnarray}
where ${\cal Y}^\ell:={}_2\!\bar Y_{\ell 0}\left(\frac{\pi}{2},0\right)$, and 
${\cal Y}^\ell_\theta$ and ${\cal Y}^\ell_{\theta\theta}$ are, respectively, the first and second derivatives of ${}_2\!\bar Y_{\ell m}(\theta,0)$ with respect to $\theta$, evaluated at $\theta=\pi/2$.
We note that ${\cal Y}^\ell$ and ${\cal Y}^\ell_{\theta\theta}$ vanish for all odd values of $\ell$, while ${\cal Y}^\ell_\theta$ vanishes for all even values of $\ell$. Inspecting Eqs.\ (\ref{JphiStat}) and (\ref{JphirStat}) we consequently find that the jumps $\J{\phi_{\ell 0}}$ and $\J{\phi'_{\ell 0}}$ are real for even $\ell$ and imaginary for odd $\ell$.

In summary, the axially symmetric piece of the IRG $\ell$-mode Hertz potential in and out of a circular geodesic orbit in Schwarzschild spacetime  ($\phi^{-}_{{-2}\ell 0}$ and $\phi^{+}_{{-2}\ell 0}$, respectively) is given analytically by Eq.\ (\ref{phiStat}), with the coefficients $C^\pm_{\ell}(r_0)$ given in Eq.\ (\ref{Cpm}) [with Eqs.\ (\ref{JphiStat}) and (\ref{JphirStat})]. The $m=0$ solution is purely real for even $\ell$ and purely imaginary for odd $\ell$. We have checked that our analytic solution (\ref{phiStat}) agrees with that obtained in Ref.\ \cite{Merlin:2016boc} using a different method, namely starting with $\Psi_4$ and using the ``angular'' inversion formula (\ref{invertion_ang_IRG}).

\section{Numerical Implementation}\label{Sec:numerics}

In this section we present an illustrative numerical implementation of our method, specializing to circular geodesic orbits in Schwarzschild spacetime. Our goal here is twofold: First, we will be able to provide some test for our formulation by comparing numerical results with results obtained analytically or using other methods. Second, we aim to illustrate the kind of numerical implementation strategy we have in mind, which, we believe, can be used to tackle more general cases. Here we do not seek sophistication in our numerical technique, and we do not concern ourselves with questions of computational performance or code optimization. We leave such matters to future work.

\subsection{Method}

Our code solves for the IRG Hertz-potential modes $\phi^{\rm IRG}_{\ell m}(r,t)$ by evolving a finite-difference version of the hyperbolic equation (\ref{Teukolsky1+1Sch}) on a fixed 1+1D grid in double-null ($u,v$) coordinates, subject to the jump conditions  (\ref{Jphi}) and (\ref{Jphiv}). This double-null approach follows the general strategy set out in Refs.\ \cite{Barack:2005nr,Barack:2007tm,Barack:2010tm}.
Our finite-difference method, detailed in Appendix \ref{FDS}, is a straightforward second-order convergent scheme. By this we mean that its {\em global} (time-accumulated) error scales as $h^2$, where $h\times h$ are the null-coordinate dimensions of a single grid cell. (To achieve this requires that the local finite-difference error at each grid cell scales with a higher power of $h$; see Appendix \ref{FDS} for details.) Higher-order convergence may be desirable in future applications. It can be achieved in a straightforward manner, following, e.g., the methods of \cite{Lousto:2005ip} or \cite{Barack:2010tm}.

The code, implemented in {\tt Mathematica}, takes as input the orbital radius $r_0$ and the mode numbers $\ell,m$, and returns the field $\phi^{\rm IRG}_{\ell m}(t,r)$. The evolution starts from initial data on two characteristic rays $v=r_*(r_0)=:v_0$ and $u=-r_*(r_0)=:u_0$, which intersect on the particle's orbit at $t=0$. The numerical integration then proceeds along successive $v={\rm const}>v_0$ rays, with the jump conditions  (\ref{Jphi}) and (\ref{Jphiv}) imposed whenever the particle's worldline [represented by the line $v=u+2r_*(r_0)$] is crossed. The future boundaries of the grid are taken at some large values of $v$ and $u$ (approximating null infinity and the event horizon, respectively), so the numerical domain has no timelike causal boundaries where boundary conditions would have been required.  For initial data we take, for simplicity, $\phi^{\rm IRG}_{\ell m}(v=v_0)=0=\phi^{\rm IRG}_{\ell m}(u=u_0)$. This choice, which violates the jump conditions at $r=r_0$, produces an initial burst of nonphysical (``junk'') radiation, which, however, dies out over time (with some $\ell$-dependent inverse-power law). The evolution proceeds until the level of junk radiation (as determined from the residual nonstationarity of the numerical solution) drops below a set threshold, and one then discards the early, junk-dominated part of the data. 

How can one be sure that the late-time solution thus obtained is the physical, ``retarded'' solution sought for? Our artificial choice of characteristic initial data means that the early part of the evolution likely contains a nonphysical component that violates the retarded boundary condition (it would contain, in particular, radiation that comes in from past null infinity and out of the past horizon). Since the numerical solution satisfies the correct jump conditions on the particle (up to numerical error), this nonphysical component may be thought of as a homogeneous (vacuum) perturbation superposed on the true, inhomogeneous physical solution. With our choice of initial conditions, we also know that this vacuum perturbation has an initial compact support. It is a well known feature of the vacuum Teukolsky equation (see, e.g., \cite{Barack:1999st}) that solutions of an initial compact support die off at late time and have no manifestation at timelike infinity. Thus, we expect the nonphysical component of the numerical solution to die off at late time, and the full solution to relax to its physical value.\footnote{In order to obtain a late-time solution other than the retarded one, one would need to adjust the form  of characteristic initial data at late retarded and advanced times, so as to represent radiation coming out of past null infinity and/or out of the past horizon. Our argument is that whenever the initial data are compactly supported, as in our implementation, the solution will relax to the retarded one at late time.}

As discussed in Sec.\ \ref{subsec:BC}, one can also check {\it a posteriori} whether the numerical solution is the physical one. For stationary ($m=0$) modes, in our IRG case,  all nonphysical solutions blow up either at infinity or at the horizon (or at both), so demonstrating boundedness of our solution (even in the limit of diminishing step size) should suffice to establish that the solution is indeed the physical one. For nonstationary ($m\ne 0$) modes, boundedness on the horizon implies that the internal solution $\phi^-$ is the physical one, though, in our IRG implementation, we have no such direct test for the external solution $\phi^+$.  

Our numerical experiments indeed suggest that the Hertz-potential field always settles down to the true, physical solution at late time (to within our controlled numerical accuracy). This is shown most convincingly by comparing with analytic results (for $m=0$) and frequency-domain numerical calculations (for $m\ne 0$), in which the correct boundary conditions were explicitly imposed. We present some of this evidence below.

\subsection{Sample results} 
 
\begin{figure}[htb]
\includegraphics[scale=0.57]{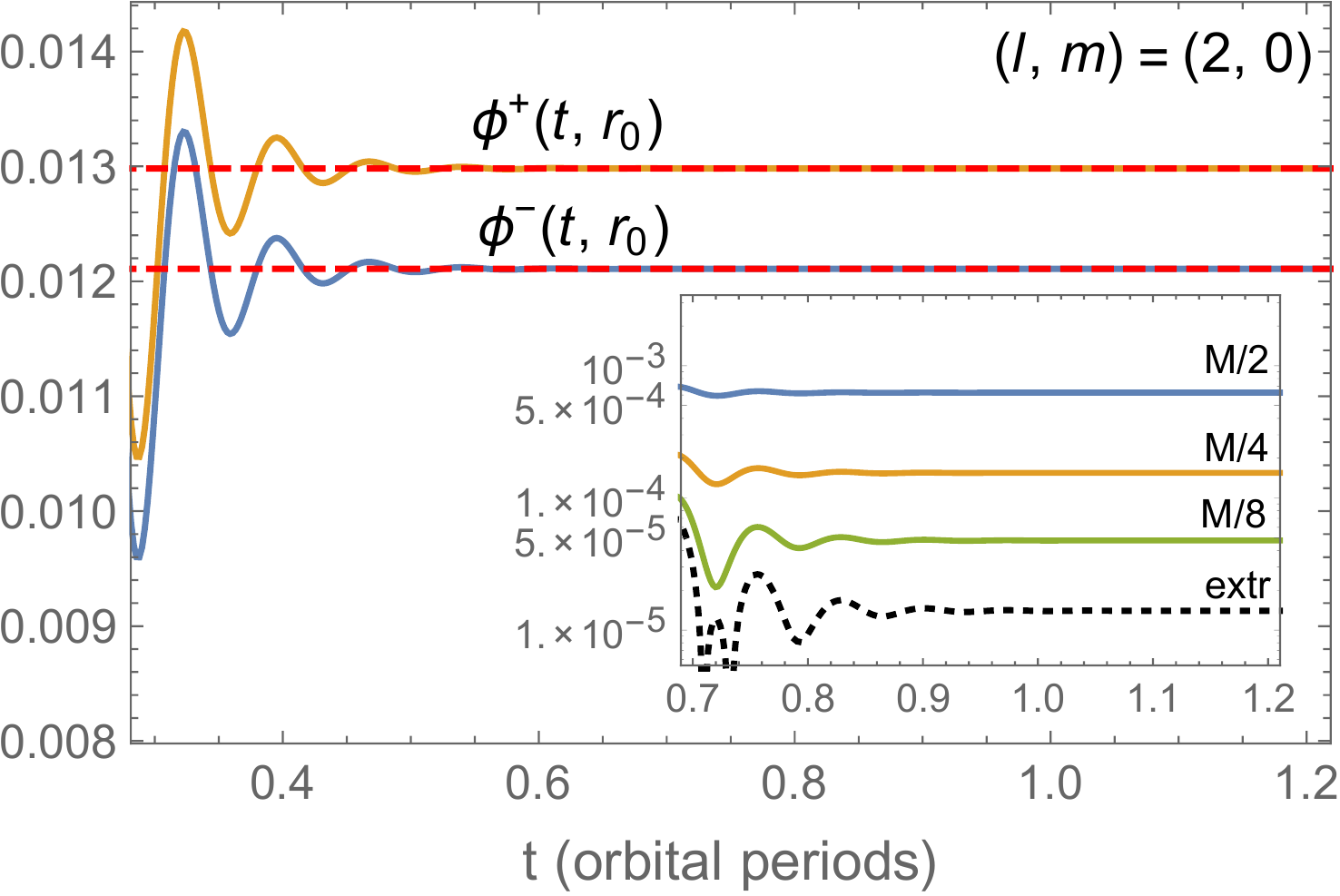}
\caption{Relaxation of the numerical solution at late time. Shown here are numerical results for $\phi^{\pm}_{20}(t,r_0)$ (divided by $\mu/M^2$), i.e.\ the $(\ell,m)=(2,0)$ mode of the IRG Hertz potential along the particle's orbit (where it is discontinuous), as a function of time. The orbit is a circular geodesic with $r_0=7M$. The early part of the solution is dominated by nonphysical junk radiation. The solution relaxes at late time to a stationary value, shown to be in agreement with that of the analytical solution (\ref{phiStat}), indicated as a dashed line. The inset shows, for $\phi^{-}_{20}$, how the agreement with the analytical solution improves with increasing numerical resolution: shown, on a semilogarithmic scale, is the magnitude of {\em relative} difference between the numerical data and the analytic value for each of $h=\{\frac{1}{2},\frac{1}{4},\frac{1}{8}\}M$, along with (in the dashed line) a Richardson extrapolation to $h\to 0$, which assumes quadratic convergence in $h$. As a rough error bar on the extrapolated value one may take the magnitude of its difference with the $h=M/8$ result, which in relative terms is $\sim 3\times 10^{-5}$.  We see that the extrapolated value agrees with the analytical result to within that error bar. }
 \label{Fig:relaxation}
\end{figure}

Figure \ref{Fig:relaxation} shows sample results for $(\ell,m)=(2,0)$, demonstrating the relaxation of the numerical solution towards the analytically known solution (\ref{phiStat}) at late time. The figure displays the behavior of the fields $\phi_{20}^+$ and $\phi_{20}^-$ on the particle (fixed on a circular geodesic orbit at $r_0=7M$), as a function of time. The inset illustrates how the asymptotic agreement with the analytical solution improves with decreasing grid size. Our highest-resolution run in this case (with $h=M/8$) took about 3 minutes on a modest laptop, outputting the correct result with a mere $\sim 5\times 10^{-5}$ relative error. If we use as a rough measure of numerical error the difference between the highest-resolution value and the value obtained via a Richardson extrapolation to $h\to 0$ (assuming quadratic convergence), we see that the extrapolated value is consistent with the analytical result to within the estimated  error.

\begin{figure}[htb]
\includegraphics[scale=0.60]{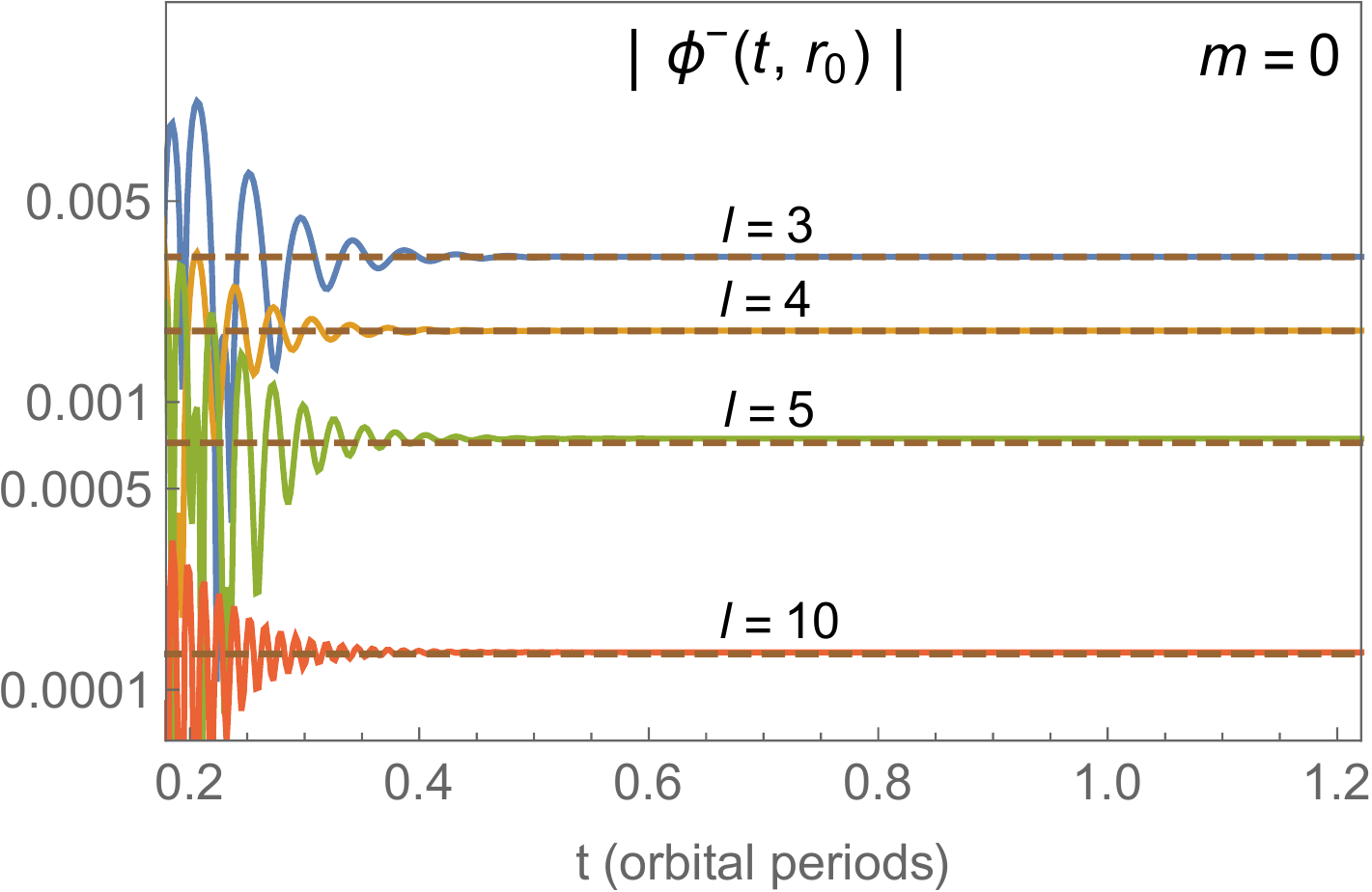}
\caption{The axisymmetric piece of the Hertz potential on the particle, for different $\ell$ values. We show here $|\phi^{-}_{\ell0}(t,r_0)|$ (divided by $\mu/M^2$) for $\ell=3,4,5,10$ and $r_0=7M$. The corresponding known analytic values are shown in dashed lines, for comparison.  Note how modes of higher $\ell$ relax faster, as expected. }
 \label{Fig:axisym_vs_t}
\end{figure}

\begin{figure}[htb]
\includegraphics[scale=0.60]{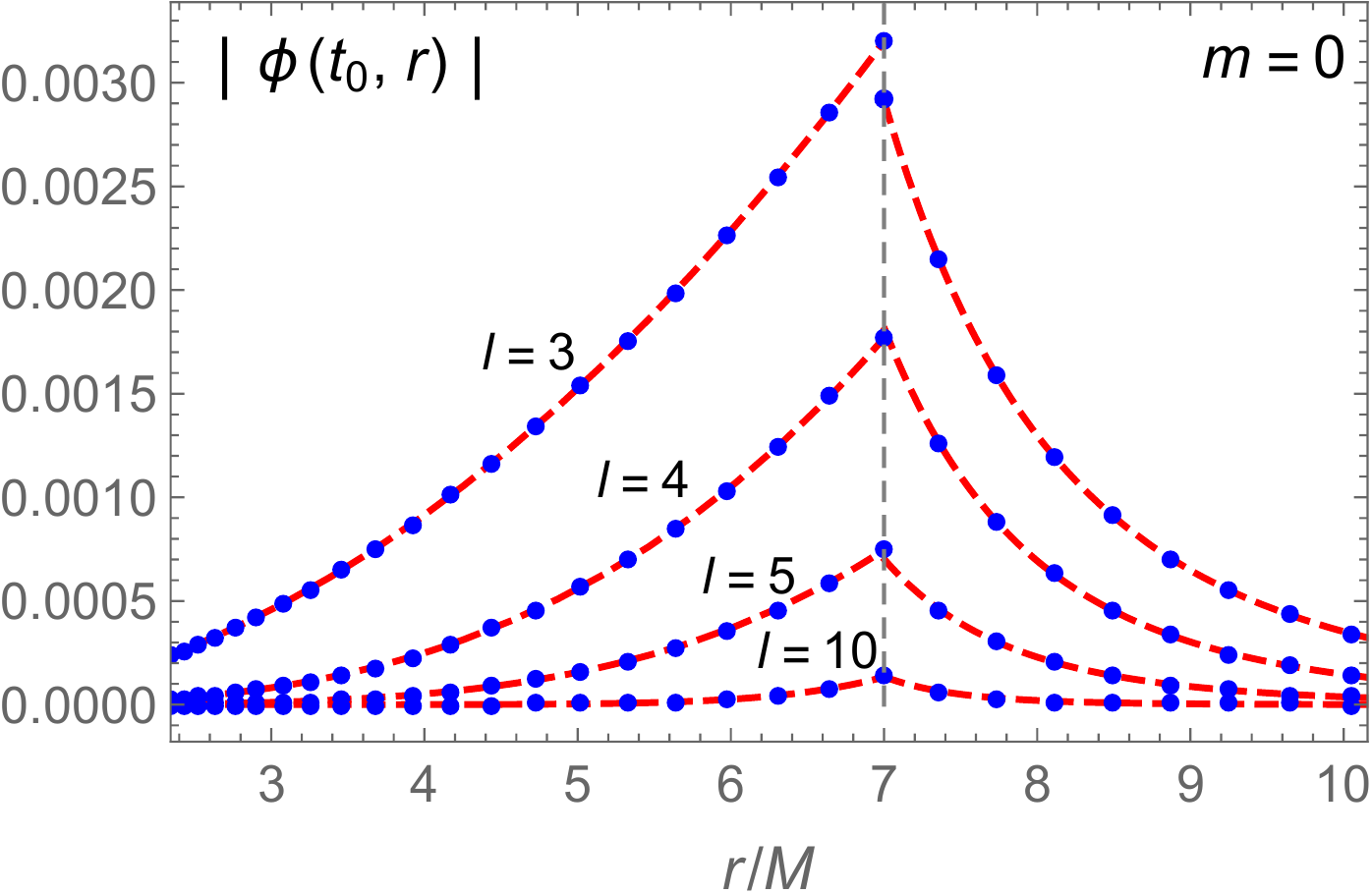}
\caption{Same as in Fig.\ \ref{Fig:axisym_vs_t}, this time showing the behavior of $|\phi^{\pm}_{\ell0}|$ as a function of $r$ at some fixed (late) time $t=t_0$. We show individual numerical data points, with the analytic solutions (dashed lines) displayed for comparison.
The fields $\phi^{-}_{\ell0}$ and $\phi^{+}_{\ell0}$ do not agree on the particle [recall Eq.\ (\ref{JphiStat})], which is clearly manifest in the $\ell=3$ case; higher-$\ell$ modes have smaller jumps, which are harder to resolve by eye in this figure.
}
 \label{Fig:axisym_vs_r}
\end{figure}

Figures \ref{Fig:axisym_vs_t} and \ref{Fig:axisym_vs_r} further test the $m=0$ solution for a variety of $\ell$ values, again comparing with the analytical solution (\ref{phiStat}). Higher multipoles exhibit smaller-scale features, the resolution of which demands smaller step sizes and hence extra computational resources. On the other hand, higher multipoles also relax faster at late time, allowing a shorter evolution, which somewhat alleviates the computational burden.

\begin{figure}[htb]
\includegraphics[scale=0.72]{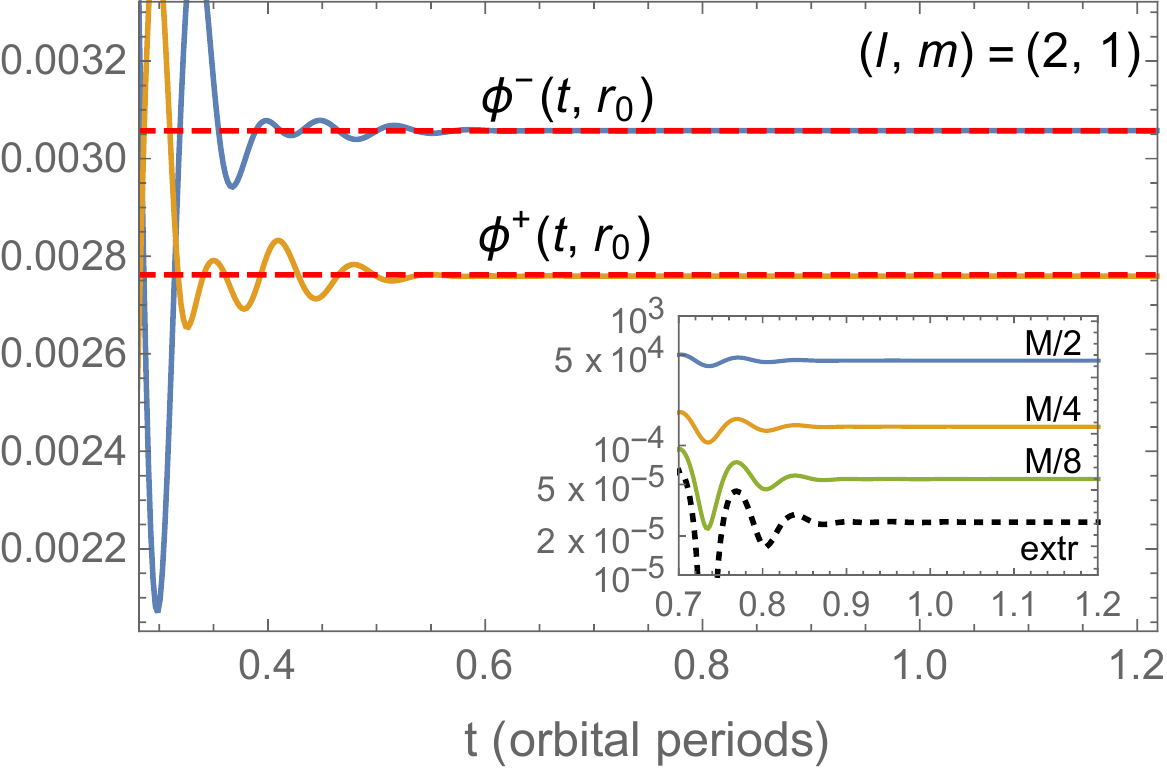}
\caption{Results for $(\ell,m)=(2,1)$. This figure is organized the same as Fig.\ \ref {Fig:relaxation}, but for convenience we show here the quantities $|\phi^{\pm}_{21}(t,r_0)\,{}_{-2}Y_{21}(\pi/2,\varphi_{\rm p}(t))|$ (divided by $\mu/M^2$), which are $t$ independent in the physical solution. Again, $r_0=7M$. Each one-sided numerical solution settles down to a constant value, which appears to be in agreement with solutions obtained using the frequency-domain approach of Ref.\ \cite{Merlin:2014qda} (dashed line). The inset explores the agreement for $\phi^{-}_{21}$ in more detail: it shows the {\it relative} difference between our numerical results and the frequency-domain ones for three choices of step size $h$, with a Richardson extrapolation to $h\to 0$ (dashed line).
As a rough error bar on the extrapolated value we take the magnitude of its difference with the $h=M/8$ result, which in relative terms is $\sim 3\times 10^{-5}$. We see that the extrapolated value agrees with the frequency-domain result to within that error bar (the numerical error of the frequency-domain value is known to be much smaller).
}
 \label{Fig:m=1_vs_t}
\end{figure}

\begin{figure}[htb]
\includegraphics[scale=0.74]{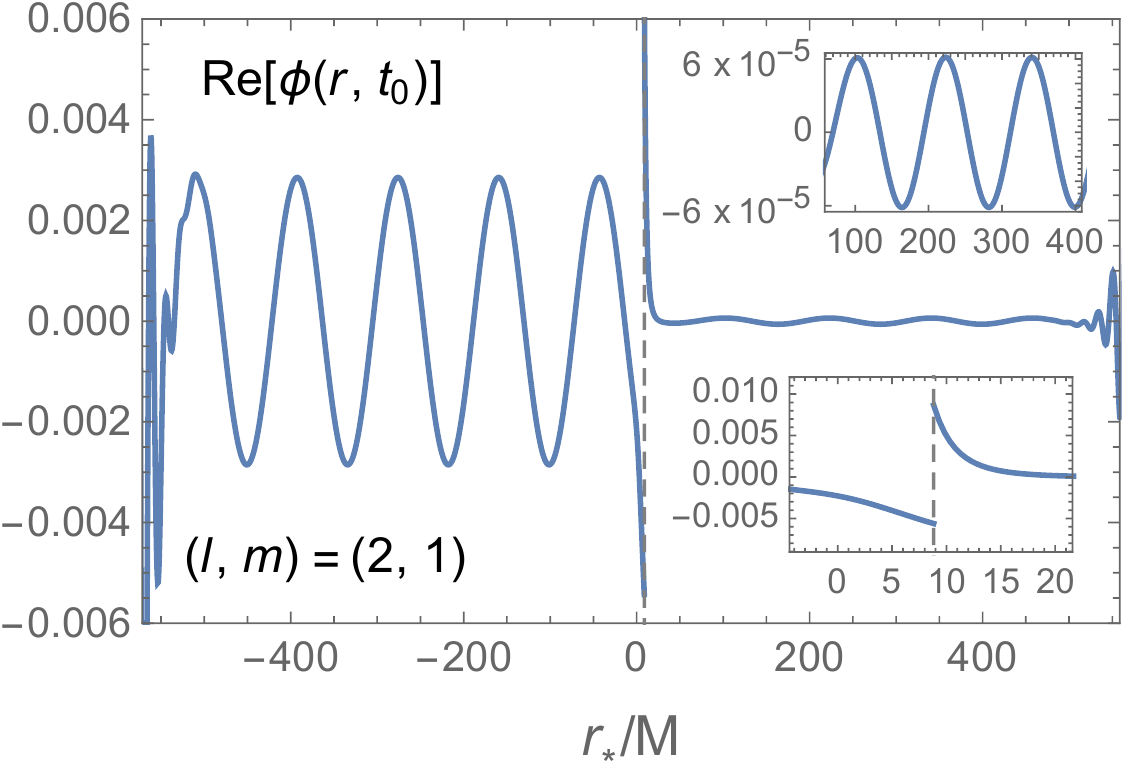}
\caption{A $t$=const slice of the same solution shown in Fig.\ \ref{Fig:m=1_vs_t}. We show the real part of the IRG field $\phi_{21}$ as a function of $r_*$ on some (late) time slice. The noisy features at both extremes are remnants of initial junk radiation, to be discarded. 
}
 \label{Fig:m=1_vs_r}
\end{figure}

Figures \ref{Fig:m=1_vs_t} and \ref{Fig:m=1_vs_r} display numerical results for a nonaxisymmetric mode of the Hertz potential ($m=1$). In this case we do not have analytical results, but we can compare with numerical solutions obtained using the frequency-domain approach of Merlin and Shah \cite{Merlin:2014qda}, who kindly provided us with numerical data generated by their code. As demonstrated in Fig.\ \ref{Fig:m=1_vs_t}, our results agree with theirs to within our (small) estimated numerical error. Figure \ref{Fig:m=1_vs_r} illustrates the behavior of the $m=1$ solution on a $t$=const slice, showing waves away from the particle and a discontinuity on it. 

\begin{figure}[htb]
\includegraphics[scale=0.82]{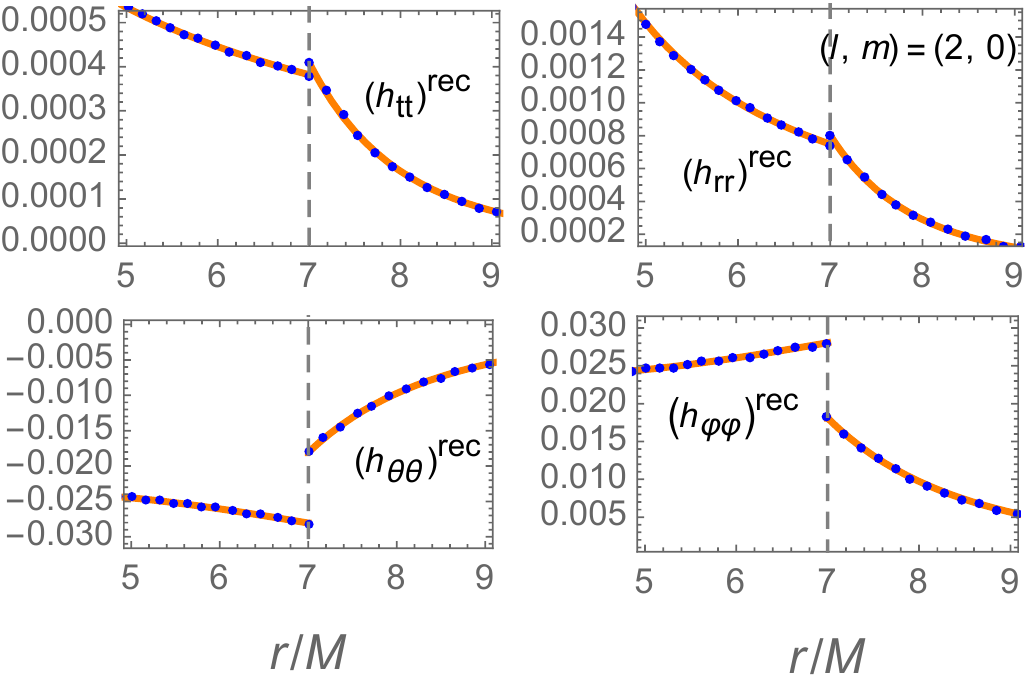}
\caption{The $(\ell,m)=(2,0)$ mode of the time-domain reconstructed IRG metric perturbation $h_{\alpha\beta}^{\rm rec\pm}$, in the vicinity of the orbit (a circular geodesic at $r_0=7M$). We show, for illustration, only four of the components (with $h_{tt}$ and $h_{rr}$ divided by $\mu/M$ and $h_{\theta\theta}$ and $h_{\varphi\varphi}$ divided by $\mu M$), on a $t$=const slice. Data points correspond to numerical data, and the solid lines come from applying the reconstruction formula (\ref{reconstruction_pm}) to the analytical solution (\ref{phiStat}). This reconstructed perturbation can be used directly as input for a self-force calculation, using the mode-sum procedure reviewed in Sec.\ \ref{subsec:mode-sum}.
}
 \label{Fig:MP}
\end{figure}

Once we have at hand the Hertz potential, reconstruction of the IRG metric perturbation becomes straightforward, through a mode-by-mode application of Eq.\ (\ref{reconstruction_pm}). Figure \ref{Fig:MP} shows, as an illustration, some of the components of the $(\ell,m)=(2,0)$ mode of reconstructed perturbation in the vicinity of the particle. The analytical solution is also shown, for comparison. We observe $h_{\alpha\beta}^{\rm rec+}\ne h_{\alpha\beta}^{\rm rec-}$ on the particle, as expected.

\section{Summary and future applications}\label{Sec:summary}

A recent body of work \cite{Keidl:2006wk,Keidl:2010pm,Pound:2013faa,Merlin:2014qda,Shah:2015nva,vandeMeent:2015lxa,vandeMeent:2016pee,Merlin:2016boc,vandeMeent:2017fqk}, reviewed in Sec.\ \ref{Sec:GSF}, showed how the gravitational self-force experienced by a point particle in orbit around a Kerr black hole can be calculated from a reconstructed metric perturbation, starting from frequency-domain solutions of the fully separated Teukolsky equations. In this approach, a suitable version of the metric perturbation is reconstructed as a sum over frequency-harmonic modes (subject to a certain regularization procedure). In the current paper we have developed a time-domain version of that approach, which circumvents the need to calculate individual frequency modes of the perturbation. Instead, each multipole ($\ell,m$ mode) of the perturbation is reconstructed (numerically) directly as a function of time. This should enable tackling some self-force problems that are not directly amenable to a frequency-domain treatment, including the very important problem of self-consistently evolving the orbit under backreaction from the self-force. 

In our new method, developed in Sec.\ \ref{Sec:scheme}, the metric perturbation is reconstructed multipole by multipole from scalarlike potentials $\phi_{\ell m}^{+}(t,r)$ and $\phi_{\ell m}^{-}(t,r)$ defined in the vacuum regions $r>r_{\rm p}(t)$ and $r<r_{\rm p}(t)$, respectively. Each of these fields is a solution of a 1+1D version of the vacuum Teukolsky equation [Eq.\ (\ref{Teukolsky1+1}), or its alternative form (\ref{Teukolsky1+1v2})], which, for $a\ne 0$, exhibits coupling between different $\ell$ modes. The crucial and most difficult step in our method is the determination of jump conditions that relate between $\phi_{\ell m}^{+}$ and $\phi_{\ell m}^{-}$ across the particle's trajectory $r=r_{\rm p}(t)$ in the 1+1D domain. A general method for doing so, for arbitrary motion in Kerr, was developed in Sec.\ \ref{Sec:jumps}. Section \ref{Sec:example} derived the jump conditions explicitly for the example of circular geodesic orbits in Schwarzschild, and Sec.\ \ref{Sec:numerics} presented a full numerical implementation for that case.

Looking ahead, let us first consider the extension of our treatment to an arbitrary motion (for now still in Schwarzschild spacetime). The most nontrivial aspect of the extension concerns the explicit determination of jump conditions for $\phi_{\ell m}^{\pm}$. For this, one needs to refer back to the general jump equations  (\ref{JumpEq1}) and (\ref{JumpEq2}), which are a coupled set of fourth-order ODEs along the orbit. Strategies for solving this set were discussed in Sec.\ \ref{Sec:jumps}. For strictly periodic orbits (including any bound geodesic orbit, even in the Kerr case), one can decompose the jumps by means of a discrete Fourier series, and solve the resulting ODEs frequency mode by frequency mode. But we see the main potential of our method in its ability to tackle nonperiodic setups, including the problem of parabolic/hyperbolic flybys or captures, and the problem of self-consistent evolution. In such cases, the jump equations may have to be solved as ODEs, with suitable initial conditions. For orbits that start or end at infinity, such conditions may most conveniently be imposed there. In the case of self-consistent evolution, initial values for the jump conditions could be well approximated by assuming exact orbital periodicity at an initial moment where the evolution is strongly adiabatic. In both cases, it remains to explore the numerical stability of the set of ODEs and develop a robust and efficient method for solving them along the orbit. 

Moving on to implementation in Kerr, the obvious additional complication comes from $\ell$-mode coupling. The 1+1D Teukolsky equation (\ref{Teukolsky1+1}) [or its alternative version (\ref{Teukolsky1+1v2})] couples between each $\ell$ mode and its nearest and next-to-nearest neighbors, so the numerical evolution problem 
would need to be recast in a matrix form, with ``all'' $\ell$ modes solved for simultaneously. In practice, a large-$\ell$ cutoff would need to be introduced, and its error controlled.  As already mentioned, we expect the narrow band-diagonal form of the matrix operator to be an advantage, computationally. We have preformed preliminary experiments evolving our 1+1D Teukolsky equation in vacuum, for Kerr, with very encouraging results. 

An additional complication, in the Kerr case, is the occurrence of coupling terms also in the jump equations (\ref{JumpEq1}) and (\ref{JumpEq2}). Here, too, the equation would need to be recast in a suitable matrix form and solved for all relevant $\ell$ modes simultaneously. We do not expect this additional hurdle to present a serious difficulty in practice.  

The method developed in this paper is, to the best of our knowledge, the only method thus far proposed capable of fully tackling the problem of time-domain calculations of metric perturbations from a point particle in Kerr spacetime. (The Lorenz-gauge approach of Dolan and Barack \cite{Dolan:2012jg} is hampered by gauge-instability problems that are yet to be resolved.) We are therefore keen to see it further developed, 
and have embarked on a program to study further applications.

\section*{Acknowledgements}

We are grateful to Cesar Merlin and Maarten van de Meent for providing us comparison data generated, upon our request, using their respective frequency-domain codes. We thank Charalampos Markakis for useful inputs during early discussions leading to this project. We gratefully acknowledge support from the European Research Council under the European Union's Seventh Framework Programme FP7/2007-2013/ERC, Grant No.\ 304978. 

\appendix

\section{Calculating the discontinuity in the Weyl scalar and its derivative}\label{AppA}

In Sec.\ \ref{Sec:example} we described the derivation of the (IRG) Hertz potential for circular geodesic orbits around a Schwarzschild black hole. Equations (\ref{Jphi}) and (\ref{Jphiv}) therein gave necessary jump conditions for the $\ell$-mode Hertz potential across the particle's orbit in the $1+1$D domain. These jumps were expressed in terms of the jumps in the corresponding $\ell$ mode of the Weyl scalar $\Psi_0$ associated with the physical perturbation---more precisely, in terms of $\J{\psi_{s=2,\ell,m}}$ and $\J{\partial_v\psi_{s=2,\ell,m}}$. In this appendix we show how the latter jumps are derived directly from the source of the $s=2$ Teukolsky equation, and provide explicit expressions for them. 

Let $\Psi_0\equiv \Psi_{s=2}$ be the Weyl scalar associated with the physical metric perturbation sourced by the point particle. We expand it in $s=2$ spherical harmonics, as in Eq.\ (\ref{expansionWeyl}):
\begin{equation} \label{expansionWeylinhom}
\Psi_{s=2}=
(r\Delta^{2})^{-1}
       \sum_{\ell=2}^{\infty}\sum_{m=-\ell}^{\ell}
       \psi_{2\ell m}(t,r) {}_2\!Y_{\ell m}(\theta,\varphi).
\end{equation}
Thus $\psi_{2\ell m}\equiv \psi^+_{2\ell m}$ for $r>r_0$ and $\psi_{2\ell m}\equiv \psi^-_{2\ell m}$ for $r<r_0$, where $\psi^\pm_{2\ell m}$ are the ``homogeneous'' $\ell$-mode fields featured in Eq.\ (\ref{expansionWeyl}). The field $\psi_{2\ell m}$ satisfies the $1+1$D $s=2$ inhomogeneous Teukolsky equation (here specialized to $a=0$ and $s=2$) with a suitable distributional source term $T$ corresponding to the energy-momentum of our geodesic pointlike particle:  
\begin{align} \label{Teukolsky1+1psi}
\psi_{,uv} + U_2(r)\psi_{,u} + V_2(r)\psi_{,v}  + W_2(r)\psi =T ,
\end{align}
where hereafter indices $s\ell m$ are dropped for brevity, and the potentials are 
\begin{equation} \label{UV2}
U_2(r)=-\frac{2M}{r^2},\quad\quad
V_2(r)=\frac{2f}{r},
\end{equation}
\begin{align} \label{W2_Sch}
W_2(r)=\frac{f}{4}\left(\frac{(\ell+3)(\ell-2)}{r^2}+\frac{6M}{r^3}\right).
\end{align}

We now need the explicit form of the $\ell$-mode source $T$, for the case of a point particle in a circular geodesic orbit. This can be worked out as explained (e.g.) in Sec.\ III.A of Ref.\ \cite{Merlin:2016boc} (where $T$ was derived for $s=-2$ and $m=0$), by using Eq.\ (A10) therein with the energy-momentum tensor (21) therein, and then decomposing into spin-2 spherical harmonics. We obtain the distributional form
\begin{align}\label{T}
T(t,r;r_0)=s_0(t,r_0)\delta(r-r_0)+s_1(t,r_0)\delta'(r-r_0)
\nonumber \\
+s_2(t,r_0)\delta''(r-r_0),
\end{align}
where the coefficients are given by 
\begin{equation}
s_n=\mu\gamma_0 \pi \tilde s_n ,
\end{equation}
with 
\begin{eqnarray} \label{s0}
\tilde s_0&=&
f_0 r_0\left[2(1-4y+10y^2) -2i\tilde m\Omega r_0(1+3y) - \tilde m^2\right] {\cal Y}(t) 
\nonumber\\
&&
-2f_0^2 r_0 \left[i\Omega r_0(1+4y) + \tilde m\right]{\cal Y}_\theta(t)
\nonumber\\
&&
- f_0^3 r_0 {\cal Y}_{\theta\theta}(t),
\end{eqnarray}
\begin{eqnarray}\label{s1}
\tilde s_1&=&
2f_0^2 r_0 \left[-2M(1+y) + i\tilde m\Omega r_0^2\right]{\cal Y}(t)
\nonumber\\
&&
+2i  f_0^3 r_0^3 \Omega{\cal Y}_\theta(t),
\end{eqnarray}
\qwe \label{s2}
\tilde s_2= M f_0^3r_0^2 {\cal Y}(t).
\ewq
We recall $y=M/r_0$ and $f_0=1-2M/r_0$, and we have further introduced here 
\begin{equation}
\tilde m:= m\gamma_0^{-2} = m(1-3M/r_0)
\end{equation}
and
\begin{equation}
{\cal Y}(t):={}_2\!\bar Y_{\ell m}\left(\frac{\pi}{2},\Omega t\right),
\end{equation}
with ${\cal Y}_\theta$ and ${\cal Y}_{\theta\theta}$ being, respectively, the first and second derivatives of ${}_2\!\bar Y_{\ell m}\left(\theta,\Omega t\right)$ with respect to $\theta$, evaluated at $\theta=\pi/2$.
We have taken $\varphi_{\rm p}(t)=\Omega t$, with $\Omega$ being the orbital angular velocity, and used $d{\cal Y}/dt=-im\Omega{\cal Y}$ and $d^2{\cal Y}/dt^2=-m^2\Omega^2{\cal Y}=-m^2(M/r_0^3){\cal Y}$.

Next, we substitute the {\it Ansatz}
\begin{align}
\psi=\psi^+(t,r)\Theta(r-r_0)+\psi^-(t,r)\Theta(r_0-r)
\\
+\psi_\delta(t,r_0)\delta(r-r_0),
\end{align}
where $\Theta(\cdot)$ is the standard Heaviside step function, into the field equation (\ref{Teukolsky1+1psi}). We note all resulting terms that are proportional to $\Theta(r-r_0)$ or to $\Theta(r_0-r)$ vanish, by virtue of $\psi^\pm(t,r)$ being homogeneous solutions. The remaining terms are supported on $r=r_0$ only, and are each proportional to $\delta$, $\delta'$ or $\delta''$. We use the distributional identities 
\begin{eqnarray}
F(r)\delta(r-r_0) &=& F(r_0)\delta(r-r_0),\nonumber\\
F(r)\delta'(r-r_0) &=&F(r_0)\delta'(r-r_0)-F'(r_0)\delta(r-r_0),\nonumber\\
F(r)\delta''(r-r_0)&=&F(r_0)\delta''(r-r_0)-2F'(r_0)\delta'(r-r_0)
\nonumber\\ 
&& +F''(r_0)\delta(r-r_0)
\end{eqnarray} 
[valid for any smooth function $F(r)$] to reexpress the coefficients of $\delta$, $\delta'$ and $\delta''$ in terms of $r_0$ (and $t$) only, and then compare the values of these coefficients across both sides of Eq.\ (\ref{Teukolsky1+1psi}), recalling the form of $T$ in Eq.\ (\ref{T}). From the coefficient of $\delta''$ one immediately obtains
\begin{equation}
\psi_\delta(t,r_0)=-4f_0^{-2}s_2(t,r_0),
\end{equation}
and subsequently comparing the coefficients of $\delta'$ and $\delta$ uniquely determines the jumps $\J{\psi}=\psi^+(r_0)-\psi^-(r_0)$ and  $\J{\psi'}=(\psi^+)'(r_0)-(\psi^-)'(r_0)$. We obtain
\begin{equation}\label{Jpsi}
\J{\psi}=
8\pi\mu\gamma_0 r_0^2\left[
\left(y^2-i\tilde m \Omega r_0\right){\cal Y}(t) 
-if_0 r_0 \Omega {\cal Y}_\theta(t)
\right],
\end{equation}
\begin{eqnarray}\label{Jpsir}
\J{\psi'}&=&
4\pi\mu\gamma_0 r_0\left\{
\left[2y^2-2-y(\lambda-4) +\tilde m^2f_0^{-1}(1+y^2\gamma_0^4)\right. \right.
\nonumber\\
&&
\left.
-2i\tilde m r_0\Omega f_0^{-1}(3-7y) \right] {\cal Y}(t)
\nonumber\\
&&
\left.
+2\left(-3i f_0r_0\Omega + \tilde m \right) {\cal Y}_{\theta}(t)
 + f_0 {\cal Y}_{\theta\theta}(t) \right\}.
\end{eqnarray}
We recall $\lambda=(\ell+2)(\ell-1)$.

In summary, and restoring all suppressed indices, Eqs.\ (\ref{Jpsi}) and (\ref{Jpsir}) give the desired jumps $\J{\psi_{s=2,\ell,m}}$ and $\J{\psi'_{s=2,\ell,m}}$ in explicit form. The jump in the $v$ derivative, needed as input for Eqs.\ (\ref{Jphi}) and (\ref{Jphiv}), is obtained via 
\begin{eqnarray} \label{Jpsiv}
\J{\partial_v\psi}&=&
\frac{1}{2}f_0\J{\psi'}+\frac{1}{2}\J{\partial_t\psi}
\nonumber\\
&=& 
\frac{1}{2}f_0\J{\psi'}-\frac{1}{2}im\Omega\J{\psi}.
\end{eqnarray}

\section{Finite-difference scheme} \label{FDS}

\begin{figure}[htb]
  \centering
\includegraphics[scale=0.47]{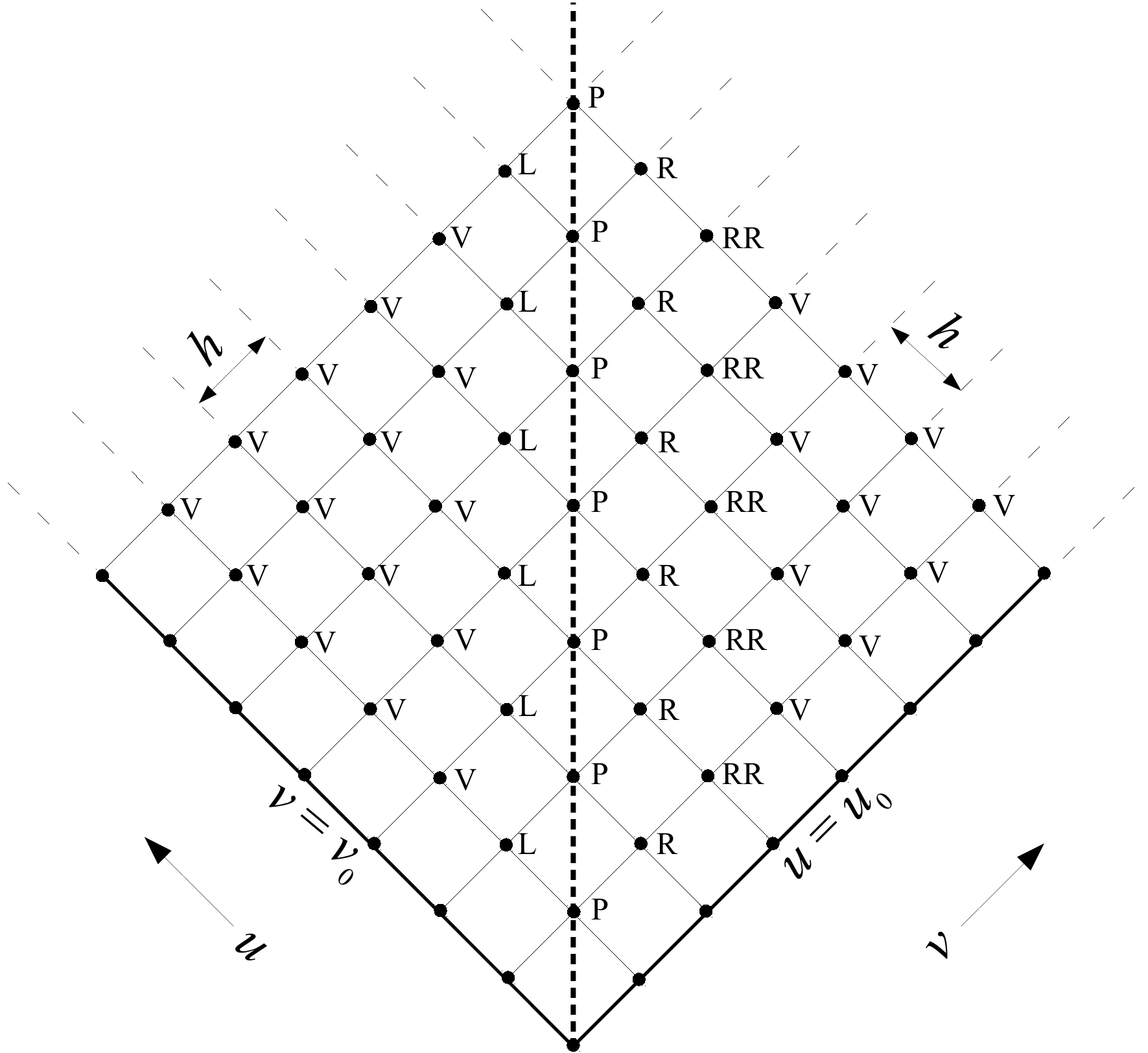}
\caption{
Our 1+1D double-null uniform grid in $(u,v)$ coordinates. The numerical integration starts with characteristic initial conditions on $v=v_0$ and $u=u_0$ and proceeds along successive lines of constant $v$.  The particle's worldline is represented by the vertical dashed line at $v-v_0=u-u_0$, and grid points are labelled in accordance with their position with respect to it. In the text we describe the finite-difference formula applied for each type of grid point. 
}
\label{Fig:grid}
\end{figure}

In this appendix we describe the finite-difference scheme used in Sec.\ \ref{Sec:numerics} for numerically computing the Hertz-potential modes in the time domain. Our treatment is based on a characteristic evolution of the vacuum field equation $(\ref{Teukolsky1+1Sch})$ on a fixed 1+1D uniform mesh in double-null $(v,u)$ coordinates. The numerical domain is depicted in Fig.\ \ref{Fig:grid}. We start with initial conditions on the rays $v=v_0$ and $u=u_0$, and evolve along successive rows (``rays'') of constant $v$.  At each step of the integration, we approximate the value of the field at a grid point using the already-known values at a few grid points in its recent ``causal past''. The form of the finite-difference formula applied depends on the position of the grid point with respect to the particle's worldline (represented by the vertical line $v-v_0=u-u_0$): For points that are sufficiently far from the worldline (those marked `V' in Fig.\ \ref{Fig:grid}) we apply a certain ``vacuum'' formula, whereas for points in the vicinity of the worldline (`L', `R' or `RR') or on it (`P')  we apply modified formulas that involve the known jumps in the value of the field and its derivatives across the worldline.  Below we shall describe the finite-difference schemes applied for each type of point.

But first we introduce some notation. We let $h$ denote the constant stepping interval (in each of $v$ and $u$), so that the $u,v$-coordinate dimensions of a single grid cell are $h\times h$. Consider an arbitrary grid point C at the coordinate position $(u,v)=(u_c,v_c)$, for some $u_c$ and $v_c$. With reference to the point C, we denote by $\phi(n,k)$ (where $n,k\in\mathbb{N}$) our finite-difference approximant for the value of the field at the grid point with coordinates $(u,v)=\left(u_c-nh,v_c- kh\right)$. Our goal is to obtain, for each such point C, the value $\phi(0,0)$ in terms of the values $\phi(n\geq 0,k\geq 0)$ (excluding $n=0=k$) that are computed in previous steps of the integration. For our second-order-convergence scheme it will prove sufficient to use the six values $\phi(1,0)$, $\phi(0,1)$, $\phi(1,1)$, $\phi(2,0)$, $\phi(0,2)$ and  $\phi(2,2)$. This requires storing field values at three successive $v$=const rays at a time. 

We note that the field value at a P-type point is ambiguous: the field there admits, in general, two different one-sided values $\phi^-$ and $\phi^+$. We have found it convenient to assign to each P point a concrete one-sided value, and in practice we take it to be the left-hand value $\phi^-$. (The choice of side here is arbitrary; assigning the value $\phi^+$ would work just the same.) This amounts to taking the orbit to pass ``just to the right'' of the line of P points. Henceforth, whenever we refer to the value $\phi(n,k)$ at a P-type point, we mean $\phi^-(n,k)$. Of course, given $\phi^-$ at a P-type point, the right-hand value may be immediately recovered using
\begin{equation}
\phi^+=\phi^- +[\phi],
\end{equation}
where $[\phi]$ is the known jump in $\phi$ at that point.

\subsection{V-type (vacuum) grid points}

V-type points are those with coordinates $(u,v)$ satisfying $\Delta v-\Delta u\geq 3h$ or $\Delta v-\Delta u\leq -2h$, where henceforth $\Delta v:=v-v_0$  and $\Delta u:=u-u_0$. Let us consider an arbitrary V-type grid point at $(u,v)=(u_c,v_c)$. To obtain a finite-difference expression for the field value at that point, it is convenient to consider the integral of both sides of the field equation (\ref{Teukolsky1+1Sch}) over the grid cell with vertices $(u_c,v_c)$, $(u_c-h,v_c)$, $(u_c,v_c-h)$ and $(u_c-h,v_c-h)$ (i.e., the grid cell with our V-type point as its upper vertex). We represent that integral with the symbol $\int_{\Diamond}$.

Let us consider Eq.\ (\ref{Teukolsky1+1Sch}) term by term. First, we have  
\begin{equation}\label{FDS:uvterm}
\int_{\Diamond} \phi_{,uv}\,dudv= \phi(0,0)-\phi(1,0)-\phi(0,1)+\phi(1,1),
\end{equation}
which is {\em exact}, and does not involve any finite-difference approximation. Next, for the fourth term in (\ref{Teukolsky1+1Sch}) we obtain 
\begin{eqnarray}\label{FDS:Wterm}
\int_{\Diamond} W(r)\phi\, dudv&=& \frac{1}{2}h^2 W(r_c)\left[\phi(1,0)+\phi(0,1)\right]
 \nonumber\\ 
&&+O(h^4),
\end{eqnarray}
where $r_c$ is the value of $r$ at the point $(u_c,v_c)$. This finite-difference approximation suffices for our purpose: since the total number of V-type grid points scales as $h^{-2}$, the accumulated error from the omitted $O(h^4)$ terms should be of $O(h^2)$ at most, consistent with the sought-for quadratic convergence. 

To obtain finite-difference approximations for the second and third terms in (\ref{Teukolsky1+1Sch}), we apply the following, more systematic procedure (following \cite{Barack:2010tm}). We first formally write $\phi$ as a double Taylor expansion in $u$ and $v$ about the point $(u_c,v_c)$, keeping terms up to $O(h^2)$:
\begin{equation}\label{Taylor}
\phi(u,v)=\sum_{i,j=0}^{2}c_{ij}(u-u_c)^i(v-v_c)^j + O(h^3).
\end{equation}
This should be valid in a vacuum neighborhood of $(u_c,v_c)$. 
Taking $(u_c,v_c)$ as our ``$(0,0)$'' point, we then apply Eq.\ (\ref{Taylor}) at each of the vacuum points $(1,0)$, $(0,1)$, $(1,1)$, $(2,0)$, $(0,2)$ and  $(2,2)$. This yields six algebraic equations for the six coefficients $c_{ij}$ ($0\leq i,j\leq 2$) in terms of the values $\phi(1,0)$, $\phi(0,1)$, $\phi(1,1)$, $\phi(2,0)$, $\phi(0,2)$ and  $\phi(2,2)$, assumed known. We solve these equations and substitute the coefficient values back into Eq.\ (\ref{Taylor}), to obtain an approximation formula for $\phi(u,v)$ valid through $O(h^2)$ in the vicinity of $(u_c,v_c)$. Using this formula with a Taylor expansion of the (smooth) potentials $U(r)$ and $V(r)$ about $r=r_c$, we obtain approximation formulas for the two terms $U(r)\phi_{,u}$ and $V(r)\phi_{,v}$, valid through $O(h)$ near $(u_c,v_c)$.
Finally, integrating over the grid cell, we obtain 
\begin{eqnarray}\label{FDS:uterm}
\int_{\Diamond} U(r)\phi_{,u} dudv&=& \frac{h}{24}\left\{
(4U+hU_{,r_*})\left[\phi(1,0)-\phi(2,0)\right] \right.
 \nonumber\\ 
&& +(28U-hU_{,r_*})\left[\phi(0,1)-\phi(1,1)\right]
\nonumber\\ 
&& \left. + 4U\left[\phi(2,2)-\phi(0,2)\right] \right\}
+O(h^4), \nonumber\\ 
\end{eqnarray}
\begin{eqnarray}\label{FDS:vterm}
\int_{\Diamond} V(r)\phi_{,v} dudv&=& \frac{h}{24}\left\{
(4V-hV_{,r_*})\left[\phi(0,1)-\phi(0,2)\right] \right.
 \nonumber\\ 
&& +(28V+hV_{r_*}) \left[\phi(1,0)-\phi(1,1)\right]
\nonumber\\ 
&& \left. + 4V\left[\phi(2,2)-\phi(2,0)\right] \right\}
+O(h^4), \nonumber\\ 
\end{eqnarray}
in which the radial functions $U$, $U_{,r_*}$, $V$ and $V_{,r_*}$ are all evaluated at $r=r_c$.

Adding together the expressions (\ref{FDS:uvterm})--(\ref{FDS:vterm}) and equating to zero [the cell integral of the right-hand side of (\ref{Teukolsky1+1Sch})], we obtain the desired finite-difference formula for $\phi(0,0)$ in the vacuum case:
\begin{equation}\label{FDS:vacuum}
\phi_{\rm Vac}(0,0) = \sum_{n,k} H_{nk}\,\phi(n,k) +O(h^4),
\end{equation}
where the only nonvanishing coefficients are 
\begin{eqnarray}
H_{10} &=& 1-\frac{h}{6}(U+7V)-\frac{h^2}{24}(U_{,r_*}+V_{,r_*}+12W),
\nonumber\\ 
H_{01} &=& 1-\frac{h}{6}(V+7U)+\frac{h^2}{24}(U_{,r_*}+V_{,r_*}-12W),
\nonumber\\
H_{11} &=& -1+\frac{7h}{6}(U+V)-\frac{h^2}{24}(U_{,r_*}-V_{,r_*}),
\nonumber\\
H_{20} &=& \frac{h}{6}(U+V)+\frac{h^2}{24}U_{,r_*},
\nonumber\\
H_{02} &=& \frac{h}{6}(U+V)-\frac{h^2}{24}V_{,r_*},
\nonumber\\
H_{22} &=& -\frac{h}{6}(U+V).
\end{eqnarray}
Here, all radial functions are evaluated at $r=r_c$.

\subsection{Near-particle grid points}

RR, R, L and P-type points are those with coordinates $(v,u)$ satisfying $\Delta v-\Delta u=+2h$, $+h$, $-h$ and $0$, respectively. For such points, the above vacuum scheme (\ref{FDS:vacuum}) does not quite work as it stands, because it involves field values at points on both sides of the particle's worldline, where the field has a jump discontinuity.

To account for the discontinuity we apply the following procedure (again following \cite{Barack:2010tm}). For a given near-particle point (RR, R, L or P) with coordinates $(u_c,v_c)$, we write down two separate Taylor expansions,
\begin{equation}\label{Taylor2}
\phi^{\pm}(u,v)=\sum_{i,j=0}^{2}c^{\pm}_{ij}(u-u_c)^i(v-v_c)^j + O(h^3),
\end{equation}
each suitable for points on the corresponding side of the particle: $\phi^{-}$ for points on the ``left'' ($r\leq r_0$) and $\phi^{+}$ for points on the ``right'' ($r> r_0$). We apply the formula to the six points $(1,0)$, $(0,1)$, $(1,1)$, $(2,0)$, $(0,2)$ and  $(2,2)$ as before. This now yields 6 equations for the 12 coefficients $c_{ij}^{\pm}$ ($0\leq i,j\leq 2$), so the system is underdetermined. However, the known jump conditions across the particle's worldline provide additional constraints: we get 6 additional relations between the $c_{ij}^{\pm}$'s by imposing the known jumps $[\phi]$, $[\phi_{,u}]$, $[\phi_{,v}]$, $[\phi_{,uu}]$, $[\phi_{,uv}]$ and $[\phi_{,vv}]$ at some point along the worldline. We choose that point to be $(0,2)$ for the RR case, $(0,1)$ for the R case, $(1,0)$ for the L case, and $(0,0)$ for the P case. Solving the resulting set of 12 equations, we obtain the 12 coefficients $c^\pm_{ij}$ in terms of the field values at the above six points and the above 6 jumps. Substituting back into (\ref{Taylor2}) again gives an approximation formula for $\phi^{\pm}$ in the vicinity of $(u_c,v_c)$, which can be used to construct approximate expressions for each of the terms in the field equation (\ref{Teukolsky1+1Sch}) near $(u_c,v_c)$. Finally, we integrate these expressions over the grid cell, using either `$+$' or `$-$' values, as appropriate (the former for cases RR and R and the latter for case L; for P points we use $\phi^+$ on the right-hand half of the cell and $\phi^-$ on its left). 

Following this procedure, and collecting all terms, we arrive at the near-particle finite-difference formula  
\begin{equation}\label{FDS:nearparticle}
\phi_{\rm X}(0,0) = \phi_{\rm Vac}(0,0) + J_{\rm X} +O(h^3),
\end{equation}
in which $X\in\{RR,R,L,P\}$, $\phi_{\rm Vac}(0,0)$ is the vacuum expression given in Eq.\ (\ref{FDS:vacuum}), and the form of the jump terms $J_{\rm X}$ depends on the point type in question:
\begin{eqnarray}
J_{\rm RR}&=&\left(\frac{1}{6}h(U+V)-\frac{1}{24}h^2 V_{,r_*}\right)[\phi]_{(0,2)},
\nonumber\\
J_{\rm R}&=&\left(1-hU-\frac{1}{2}h^2 W +\frac{1}{24}h^2 U_{,r_*}\right)[\phi]_{(0,1)}
\nonumber\\
&& -\frac{1}{6}h^2(U+V)[\phi_{,v}]_{(0,1)},
\nonumber\\
J_{\rm L}&=&-\left(\frac{1}{6}h(U+V)+\frac{1}{24}h^2U_{,r_*}\right)[\phi]_{(1,0)}
\nonumber\\
&& +\frac{1}{6}h^2(U+V)[\phi_{,u}]_{(1,0)} ,
\nonumber\\
J_{\rm P}&=& -\left(1-hV-\frac{1}{24}h^2V_{,r_*}\right)[\phi]_{(0,0)}
-\frac{1}{2}h^2 V[\phi_{,v}]_{(0,0)}
\nonumber\\
&&+\left(h-\frac{1}{3}h^2 U -\frac{5}{6}h^2V\right)[\phi_{,u}]_{(0,0)}
\nonumber\\
&& -\frac{1}{2}h^2\left([\phi_{,uv}]_{(0,0)}+[\phi_{,uu}]_{(0,0)}\right).
\end{eqnarray}
Here, all radial functions are evaluated at the point $(0,0)$ under consideration, while the jumps are evaluated at the appropriate worldline point, indicated in subscript. Note that we have truncated the expressions at $O(h^2)$, leaving an $O(h^3)$ error in Eq.\ (\ref{FDS:nearparticle}). We allow ourselves to do so because the total number of near-particle (RR, R, L and P-type) points scales only like $1/h$, so the accumulated error from the omitted $O(h^3)$ local term should scale as $h^2$ at most, consistent with quadratic convergence. 
We also note that $J_{\rm P}$ may be written, through the relevant $O(h^2)$, in the alternative form
\begin{eqnarray}
J_{\rm P}&=& 
-\frac{1}{2}\left([\phi]_{(0,0)}+[\phi]_{(1,1)}\right)
-\frac{1}{4}h\left([\phi_{,r_*}]_{(0,0)}+[\phi_{,r_*}]_{(1,1)}\right)
\nonumber\\
&&
+\left(hV+\frac{1}{24}h^2V_{,r_*}\right)[\phi]_{(0,0)}
-\frac{1}{2}h^2 V[\phi_{,v}]_{(0,0)}
\nonumber\\
&&-\left(\frac{1}{3}h^2 U +\frac{5}{6}h^2V\right)[\phi_{,u}]_{(0,0)},
\end{eqnarray}
which involves jumps in the field and its first derivatives only [at the expense of requiring the jumps at the point $(1,1)$ as well]. This form may be simpler to use in practice. 

Equations (\ref{FDS:vacuum}) and (\ref{FDS:nearparticle}) constitute our finite-difference scheme. 
We see that, at the required order, the implementation of our scheme requires knowledge of the jumps in the field and its first derivatives only. Jumps in higher derivatives may be required for scheme with convergence faster than quadratic.

\raggedright
\bibliography{mybib}

\end{document}